\documentclass[superscriptaddress,secnumarabic,amssymb,amsmath,nobibnotes,aps,prd,showkeys,showpacs,twocolumn,nofootinbib]{revtex4-1}%

\usepackage[OT1]{fontenc}
\usepackage[utf8]{inputenc}
\usepackage{bm}
\usepackage{mathrsfs}
\usepackage[all]{xy}
\usepackage{epsfig,subfigure}
\usepackage{amssymb, amsmath, amsfonts}
\usepackage[english]{babel}
\usepackage{makeidx}
\usepackage{mathrsfs}
\usepackage[colorlinks=true,breaklinks=true]{hyperref}
\usepackage{graphicx, wrapfig, epsf}
\usepackage{float}
\usepackage{comment}
\usepackage{ulem}
\usepackage[dvipsnames, table]{xcolor}
\hypersetup{urlcolor=BlueViolet,
	        citecolor=Plum,
	        linkcolor=OliveGreen}

\definecolor{Gray}{rgb}{.9,.9,.9}

\begin{document}

\title{Propagation and lensing of gravitational waves in Palatini \texorpdfstring {$f(\hat R)$ gravity}{Lg}}

\author{Sreekanth Harikumar}
\email{sreekanth.harikumar@ncbj.gov.pl}
\thanks{corresponding author}
\affiliation{National Centre for Nuclear Research, Ludwika Pasteura 7, Warsaw 02-093, Poland
}

\author{Laur J\"arv}
\email{laur.jarv@ut.ee}
\affiliation{Laboratory of Theoretical Physics, Institute of Physics, University of Tartu,
W. Ostwaldi 1, 50411 Tartu, Estonia
}

\author{Margus Saal}
\email{margus.saal@ut.ee}
\affiliation{Laboratory of Theoretical Physics, Institute of Physics, University of Tartu,
W. Ostwaldi 1, 50411 Tartu, Estonia
}

\author{Aneta Wojnar}
\email{awojnar@ucm.es}
\affiliation{Department of Theoretical Physics \& IPARCOS, Complutense University of Madrid, E-28040, 
Madrid, Spain}

\author{Marek Biesiada}
\email{marek.biesiada@ncbj.gov.pl}
\affiliation{National Centre for Nuclear Research, Ludwika Pasteura 7, Warsaw 02-093, Poland
}

\begin{abstract}
Accelerated expansion of the Universe prompted searches of modified gravity theory beyond general relativity, instead of adding a mysterious dark energy component with exotic physical properties. One such alternative gravity approach is metric-affine Palatini $f(\hat{R})$ theory. By now routine gravitational wave detections have opened a promising avenue of searching for modified gravity effects. Future expected cases of strong lensing of gravitational waves will enhance this opportunity further. 
In this paper, we present a systematic study of the propagation and gravitational lensing of gravitational waves in Palatini $f(\hat R)$ gravity and compare it with general relativity. Using the WKB approximation we explore the geometric-optical limit of lensing and derive the corrections to the measured luminosity distance of the gravitational source. In addition, we study the lensing by the Singular Isothermal Sphere lens model and show that Palatini $f(\hat{R})$ modifies the lensing potential and hence the deflection angle. Then we show that the lens model and chosen theory of gravity influences influence the rotation of the gravitational wave polarization plane through the deflection angle. To be more specific we discuss the $f(\hat R)=\hat R+\alpha \hat R^2$ gravity theory and find that the modifications comparing compared to general relativity are negligible if the upper bound of $\alpha \sim 10^{9} \, $m$^2$ suggested in the literature is adopted. However, this bound is not firmly established and can be updated in the future.  
Therefore, the results we obtained could be valuable for further metric-affine gravity vs. general relativity tests involving lensing of gravitational waves and comparison of luminosity distances measured from electromagnetic and gravitational wave sources.

\end{abstract}
\maketitle

\section{Introduction}
The first detections \cite{LIGOScientific:2016aoc} followed by ongoing routine observations \cite{LIGOScientific:2021djp} of gravitational waves (GW) opened a new range of possibilities not only to investigate astrophysical phenomena inherently hidden to the electromagnetic wave domain but also to test the theory of gravity. In addition, we have our first direction detection of GW signal from the Pulsar Timing Array \cite{Antoniadis:2023rey, NANOGrav:2023gor, Xu:2023wog, Reardon:2023gzh}. So far the results obtained only strengthened our confidence in general relativity (GR) and severely constrained some of the alternative gravity theories
\cite{scalartensor,dead_ends, Tessa_test}. One of the predictions of the GR is light bending by massive objects. This phenomenon underlies the theory of gravitational lensing \cite{Gravitational_lenses1992}. Hence, in the era of GW astronomy, the detection of GW signals lensed by massive sources along the line of sight is highly anticipated.  
So far, however, there is no observational evidence for the lensed GW signals in LVK data   \cite{Hannuksela:2019kle, LIGOScientific:2021izm, LIGOScientific:2023bwz}. 

The detection of lensed GW signals will open up unique opportunities for precision cosmology \cite{Sereno:2011ty, Cao:2019kgn, Liao:2017ioi, Li:2019rns, Hannuksela:2019kle, Biesiada:2021pzo, Grespan:2023cpa}, detection of Intermediate Mass Black Holes (IMBH)\cite{Lai:2018rto,meena2023gravitational}, %stellar mass objects \cite{Diego:2019rzc}, 
as well as detection of low mass halos and Primordial Black Holes \cite{Diego:2019rzc, Oguri:2020ldf}. It can also lead to the tests of GR \cite{Ezquiaga:2020dao,Goyal:2020bkm, Fan:2016swi, Collett:2016dey,Sharma:2023vme,Mishra:2023vzo} and other fundamental interactions \cite{Baker:2016reh}. Next generation of GW detectors like the Einstein Telescope (ET) \cite{Maggiore:2019uih}, Cosmic Explorer (CE) \cite{Evans:2021gyd} and the space-based detectors such as DECIGO \cite{Kawamura:2020pcg}, LISA \cite{lisa_2017}, Taiji \cite{Luo:2021qji} and TianQin \cite{TianQin:2015yph} are designed to observe distant sources up to redshifts $z \sim 20$ and higher. Some of these sources located at such large cosmological distances should be lensed. Currently, the literature discussing the predictions for gravitational wave lensing rate is rich. For example, current generation detectors are expected to detect 1 lensed signal per year for A+ \cite{Li:2018prc}. On the other hand, ET has a much higher estimate of expected 50-100 strongly lensed events per year \cite{Biesiada:2014kwa, Yang_2019, Lilan2021} which is not surprising as the accessible volume will be three orders of magnitude larger than in the current generation detectors. Predictions for the DECIGO are a bit less optimistic due to contamination by unresolved sources, yet 50 strongly lensed BH-BH systems should be detected each year of DECIGO operation \cite{Piorkowska2021}.

In the much more familiar context of electromagnetic lensing the approximation of geometric optics (GO) is used, which is justified by the fact that the wavelength is much smaller in comparison to the typical size of lens objects (stars or galaxies). 
Regarding GWs, the frequency range currently probed by ground-based detectors covers \mbox{10~Hz $<f<10$ kHz}. Future space-borne detectors, will probe \mbox{0.1 mHz $<f<100$ mHz} -- LISA and \mbox{1 mHz $<f<100$ Hz} -- DECIGO. This corresponds to the GW wavelengths of \mbox{$10^4$ m $< \lambda <10^7$ m} in ground-based and \mbox{$10^6$ m $< \lambda <10^{12}$ m} in space-borne detectors. In the case of lenses whose Schwarzschild radii are comparable to $\lambda$ wave optics (WO) approach is crucial \cite{Nakamura:1999uwi, Takahashi:2003ix}. Hence, for ground-based detectors, WO should be used for the lenses less massive than $10^4\; M_{\odot}$, while in the case of space-borne detectors, this upper mass limit reaches $10^9\; M_{\odot}$.

Even though propagation and detection of GW's involve the weak field regime, the structure of the waveform is created in violent processes in a strong gravity regime. Hence, the analysis of the waveforms offers a unique opportunity to test competing theories of gravity. GR predicts only two polarization modes known as the `\textit{plus}' and `\textit{cross}'. However,
other viable metric theories of gravity typically predict more than these two polarizations for a generic gravitational wave. In fact, the most general weak gravitational wave is composed of six modes of polarization, expressible in terms of the six ``electric'' components of the Riemann tensor \cite{Will_book, Eardley73, Thorne73}. Therefore the detection of extra polarizations has been considered as a smoking gun for modified gravity. Yet, it has been shown that even in the GR strong gravitational lensing may distort the GW waveform in in such a way that extra polarizations appear \cite{Hou:2019wdg, Dalang:2021qhu, Cusin:2019rmt}. This is another motivation for studying GW lensing in alternative theories of gravity. In what follows, we are going to examine the effect of modified gravity on the lensed gravitational signal. To our knowledge, this is the first work dealing with such a problem.

Here, we focus upon Palatini $f(\hat{R})$ gravity \cite{Olmo:2011uz}, which is one of the simplest extensions of GR. The theory does not introduce additional degrees of freedom, and there are no instabilities of the kind found in the metric $f(R)$  theories \cite{Sotiriou:2006sf}. Palatini formulation has implications for cosmology \cite{Stachowski:2016zio, Szydlowski:2017uuy, TeppaPannia:2018ale, Pinto:2018rfg, Camera:2022myt, Gialamas:2023flv} and astrophysical objects \cite{Olmo:2019qsj, Olmo:2019flu, Wojnar:2020txr, Wojnar:2020frr, Benito:2021ywe, Wojnar:2021xbr, Kozak:2021ghd, Sarmah:2021ule, Kalita:2022trq, Kozak:2023axy}, 
however in the vacuum and radiation dominated regions the theory reduces to GR plus an effective cosmological constant, 
and easily passes the Solar System tests \cite{Toniato:2019rrd, Bonino:2020wps, Hernandez-Arboleda:2023abv}. Gravitational waves in Palatini $f(\hat{R})$ carry only two polarizations, like in the GR \cite{Alves:2009eg}. Therefore if one is interested in testing the Palatini $f(\hat{R})$ gravity with GWs, strong lensing effects are of particular interest.

The paper is structured as follows. Section \ref{pal} introduces the essential features of Palatini $f(\hat R)$ gravity necessary for further discussion. In the section \ref{go}, we systematically derive the geometric optics limit in GR and in Palatini $f(\hat R)$ gravity. Then in section \ref{GWampl} the propagation of GWs in both theories is studied. In Section \ref{sectionV} we study the rotation of the GW  polarization plane in  GR and compare it with Palatini f(R) gravity. Then as an example of lensing, we discuss Singular Isothermal Sphere (SIS), which is a reliable model for a galaxy scale lens,  in the context of Palatini $f(\hat{R})$  gravity. Differences between GR and Palatini formulation are revealed and we conclude the paper in section \ref{conclusion}.

\section{Palatini \texorpdfstring{$f(\hat{R})$ gravity}{Lg}}\label{pal}

Let us begin by briefly recalling some of the key features of Palatini $f(\hat R)$ gravity. The action of the theory can be written down as \cite{Olmo:2011uz}
\begin{equation}
S[g,\hat\Gamma,\psi_m]=\frac{1}{2\kappa^2}\int d^4x \sqrt{|g|}f(\hat{R}) +S_{\text{matter}}[g,\psi_m],\label{action}
\end{equation}
where $\kappa^2=8\pi G/c^4$, $\hat{R} \equiv g^{\mu\nu}\hat{R}_{\mu\nu}(\hat\Gamma)$ is the Palatini curvature scalar, and $|g|$ the absolute value of the determinant of the metric tensor. In contrast with the usual metric approach, that is, instead of assuming that the connection is the Levi-Civita one of the metric $g$, in the Palatini formalism one considers an arbitrary torsion-free (symmetric in lower indices) connection $\hat\Gamma$ which is then used to construct the Riemann and Ricci curvatures. The curvature scalar $\hat R$ is built from both geometric structures, i.e., the metric and the independent of it, connection. On the other hand, the matter action depends on the metric $g$ and matter fields $\psi_m$ only,\footnote{That is, the matter fields are coupled to the metric only. One may also couple them to the connection \cite{Iosifidis:2021nra}, this broader class is usually called metric-affine gravity.} which guarantees that the matter energy-momentum tensor $T_{\mu\nu}=-\frac{2}{\sqrt{|g|}}\frac{\delta S_m}{\delta g_{\mu\nu}}$ obeys the usual conservation law \cite{Koivisto:2005yk},
\begin{align}
\label{eq: matter conservation}
    \nabla_\mu T^{\mu \nu} = 0,
\end{align}
with respect to the Levi-Civita covariant derivative $\nabla$ of the metric $g$.
 
The variation of the action \eqref{action} is taken with respect to both structures; the metric one gives 
\begin{equation} 
f'(\hat{R})\hat{R}_{\mu\nu}-\frac{1}{2}f(\hat{R})g_{\mu\nu}=\kappa^2 T_{\mu\nu}\label{structural},
\end{equation}
where prime is understood here as differentiating the function $f$ with respect to its argument i.e. the curvature scalar $\hat{R}$. Contracting the above equation with the metric $g_{\mu\nu}$ provides an algebraic relation between the Palatini curvature scalar and the trace of the energy-momentum tensor $T\equiv g^{\mu\nu}T_{\mu\nu}$,
\begin{equation}\label{trace}
    f'(\hat{R})\hat{R}-2f(\hat{R})=\kappa^2 T.
\end{equation}
This feature allows to solve the above equations in some particular choices of the function $f(\hat{R})$, providing 
$\hat R=\hat{R}(T)$.

On the other hand, the relation between the connection and the metric tensor is given by the variation of (\ref{action}) with respect to $\hat\Gamma$, which can be written as
\begin{equation} 
\hat\nabla_\beta(\sqrt{|g|}f'(\hat{R}(T))g^{\mu\nu})=0.\label{con}
\end{equation}
The above covariant derivative is understood as the one defined by the independent connection $\hat\Gamma$. Defining a new metric tensor 
\begin{equation}\label{confmetric}
    \hat{g}_{\mu\nu} = f'(\hat{R}(T))g_{\mu\nu}
\end{equation}
which is conformally related to the original metric $g_{\mu\nu}$, the equation \eqref{con} can be expressed as
\begin{equation}
\hat{\nabla}_\beta(\sqrt{|\hat{g}|}\hat{g}^{\mu\nu})=0.\label{con22}
\end{equation}
Therefore, the connection $\hat\Gamma$ happens to be the Levi-Civita one with respect to the conformal metric $\hat g$ \eqref{confmetric}. It is related to the Levi-Civita connection $\Gamma$ of the original metric $g$ by
\begin{equation}
\hat{\Gamma}^{\alpha}_{\mu\nu} =  \Gamma^{\alpha}_{\mu\nu} + \frac{1}{2f'}[ \delta ^{\alpha}_{\nu} \partial_{\mu}f' + \delta ^{\alpha}_{\mu}\partial_{\nu}f' - g_{\mu\nu}\partial^{\alpha}f'].
\end{equation}
As $\hat\Gamma$ can be computed from the original metric and the trace of the matter energy-momentum, it may be considered an auxiliary field that can be integrated out, with the implication that the degrees of freedom are related to the metric tensor $g$ only. Therefore, in the end, the connection is not an independent dynamical quantity in the theory. 

Equivalently, we can apply the conformal transformation \eqref{confmetric} to the metric field equation \eqref{structural} to encapsulate the gravitational dynamics in terms of the conformal metric $\hat{g}$,
\begin{equation}\label{P_scalartensor}
\hat{G}_{\mu\nu} = \kappa^2 \hat{T}_{\mu\nu} -\frac{1}{2}\hat{g}_{\mu\nu}\hat{U}(f') \,,
\end{equation}
where  $\hat{G}_{\mu\nu}$ is the Einstein tensor computed from the Levi-Civita connection $\hat{\Gamma}$ of $\hat{g}$, while $\hat{T}_{\mu\nu}=(f')^{-1}{T}_{\mu\nu}$ is the conformally transformed matter energy-momentum tensor, and the effective ``potential'' is given by
\begin{equation}
    \hat{U} = \frac{\hat{R} f'(\hat{R})-f(\hat{R})}{f'(\hat{R})^2} \,.
\end{equation}
The conformal energy-momentum does not obey the conformal conservation law, but\footnote{To see the relation between the geometric objects in different frames, see e.g.\ \cite{Stachowski:2016zio}.}
\begin{align}
\label{eq: matter nonconservation}
  \hat{\nabla}_\mu \hat{T}^{\mu\nu} &= \frac{1}{2\kappa^2} \hat{\nabla}^\nu \hat{U}(f') \,.  
\end{align}

Let us notice that if $f'(\hat R(T))$ is a constant, the connection in \eqref{con22} boils down to the Levi-Civita connection of the original metric $g$. This happens when one deals with trace-free  matter energy-momentum, $T=0$, or with a linear lagrangian. For the vanishing $T$ the field equations \eqref{structural} and \eqref{P_scalartensor} coincide and reduce to \cite{ferraris1993universal,ferraris1994universality}
\begin{equation}
   \hat R_{\mu\nu} - \Lambda(\hat R_0)g_{\mu\nu}=0,
\end{equation}
where the ``potential'' $\hat{U}$ acts like an effective cosmological constant that depends on the constant vacuum value of $\hat R$, that is
\begin{equation}\label{lambda}
   \Lambda(\hat R_0)\equiv f(\hat R_0)/2f'(\hat R_0) 
\end{equation}
with $\hat R_0=\hat R(T=0)$. Although its value does depend on a given choice of the $f(\hat R)$ model, it always leads to a unique %exterior 
solution for each choice of $f(\hat R)$. Therefore, in the vacuum or radiation environment (for which also $T=0$), the Palatini $f(\hat R)$ gravity reduces to GR with a cosmological constant. One can argue that the theory passes Solar System tests because of this property. Indeed, as examined in detail for analytic functions $f(\hat R)$ \cite{Toniato:2019rrd}, the center-of-mass orbits are the same as in GR, while modifications appearing in the terms related to the momentum and energy in the Euler equation are not sensitive enough for the current experiments which use the Solar System orbits.\footnote{However, let us mention that these considerations assume Minkowski background. Moreover, when atomic level experiments
are in our reach \cite{schwartz2019post, olmo2008hydrogen, olmo2007violation, Kozak:2021ghd, Wojnar:2022dvo, Kozak:2023axy, Hernandez-Arboleda:2023abv}, 
one may arrive at the required accuracy to confront the theory against the data provided by the experiments performed in the Solar System.} 

For the purposes of the GW physics one can neglect the cosmological constant in GR \cite{naf2009gravitational}. In this paper, we assume that the Palatini $f(\hat{R})$ effective cosmological constant $\Lambda(\hat{R}_0)$ is compliant with the observations of our universe and we can drop it from the GW calculations, although in general
there might exist specific choices of $f(\hat R)$ for which the quantity \eqref{lambda} is not negligible.

In the context of $f(\hat R)=\hat R+\alpha \hat R^2$ gravity
it has been observed that the value of the parameter $\alpha$ is connected to the curvature regime \cite{Olmo:2005zr}. This relationship arises from the fact that the Palatini curvature scalar is directly proportional to the trace of the energy-momentum tensor, resulting in a similar dependence for its value. Furthermore, when examining the weak-field limit analytically, it has been found that $|\alpha|$ is approximately less than $2\times 10^{8}\rm m^2$ \cite{Olmo:2005zr}. However, due to uncertainties in microphysics, the experiments conducted within the Solar System have not been able to establish any constraints on these parameters \cite{Toniato:2019rrd}. On the other hand, taking microphysical properties into account, seismic data from Earth constrained the theory's parameter to $|\alpha|\lesssim 10^9 \text{m}^2$ \cite{Kozak:2023axy, Kozak:2023ruu} while analyzing neutron stars' equations of state and combining the results with the observational data put the bound to be $|\alpha| \lesssim 10^6$ m$^2$ \cite{Lope-Oter:2023urz}. 
The lower bound of the parameter $\alpha \gtrapprox -7\times 10^7 \text{ m}^2$ was obtained in \cite{Wojnar:2023bvv} ensuring microscopic stability of matter.
Note that in the case of studies in the non-relativistic limit, it was shown that only the quadratic term is relevant; further terms entering the equations are of the sixth or higher order \cite{Toniato:2019rrd}, see the Poisson equation in the section \ref{GWampl}. Similar to the case of general relativity, none of the $f(\hat R)$ models can adequately explain the rotation curves of galaxies \cite{Hernandez-Arboleda:2022rim, Hernandez-Arboleda:2023abv}. As a result, no constraints have been derived from galaxy catalogs thus far. 
In contrast, the latest cosmological data such SNIa and BAO data provides bounds on the parameter $\alpha$ about forty orders of magnitude bigger, that is, $|\alpha| \leq 10^{49}\,\text{m}^2$ \cite{Gomes:2023xzk}.

In our investigation, the most relevant feature of the family of the Palatini theories is that the gravitational waves are propagating perturbations of the background conformal metric $\hat g$. It is so since in the case of the absence of the anisotropic stress, the tensor perturbations for the conformal and the spacetime metrics coincide \cite{jimenez2015tensor,jimenez2017gravitational,jimenez2018born}. We will use that fact further in this paper.

\section{Propagation of gravitational wave in the geometric optics regime}\label{go}

In order to be self-consistent and not to refer the reader to dispersed pieces of literature, we present in this section the basic formalism of GW propagation in the eikonal approximation. We start with the GR and then extend our analysis to the Palatini gravity. Such an approach allows us to trace similarities and differences between these theories.

\subsection{General relativity}

In the study of gravitational waves over some arbitrary background, the metric describing the background geometry (of larger amplitude) is considered to be smooth over a typical characteristic length scale $L_{B}$.
Gravitational waves constitute perturbations of much smaller amplitude propagating on top of the larger ones. The typical scale of spatial variation of this perturbation is denoted by $\ell$, which obeys the condition $\ell \ll L_{B} $. Therefore, the configuration can be described as 
\begin{equation} \label{g and h}
    g^{\text{total}}_{\mu\nu} = g_{\mu\nu} +  {h}_{\mu\nu},
\end{equation}
where 
a natural distinction can be made in frequency space with $g_{\mu\nu}$ being a slowly varying background (or a low momenta component) while the gravitational wave described by $ {h}_{\mu\nu}$ is a high frequency fluctuation. How fast the metric components vary is given by how their derivatives change, that is, $\partial g \sim g/L_{B}$ and $\partial h \sim h/\ell$. The separation of scales in frequency space is given by $\mathfrak{f} \gg \mathfrak{f}_{B} $. The method of separation of scales and the expansion based on that is called the \textit{short-wave expansion} \cite{Maggiore:2007ulw}. Following this procedure, we expand the Ricci tensor as 
\begin{equation}\label{Rmn expand}
    R^{\text{total}}_{\mu\nu} = R_{\mu\nu} + R_{\mu\nu}^{[1]} +R_{\mu\nu}^{[2]} + \ldots
\end{equation}
where $R_{\mu\nu}$ is the background Ricci tensor containing only the slowly varying component of the  metric,
while on the other hand,
$R_{\mu\nu}^{[1]}$ and $R_{\mu\nu}^{[2]}$ contain terms which are linear and quadratic in the metric perturbation ${h}_{\mu\nu}$, respectively. As $R_{\mu\nu}^{[1]}$ is linear, it contains only higher frequency modes whereas $R_{\mu\nu}^{[2]}$ is quadratic and contains both high and low frequency modes. 

We can rewrite Einstein's equations as
\begin{equation} \label{EE}
      R^{\text{total}}_{\mu\nu} = \kappa^2 \left (T_{\mu\nu} -\frac{1}{2} g_{\mu\nu} T, \right)
\end{equation}
and the distinction of scales guarantees us the decomposition into two parts \cite{Maggiore:2007ulw}:
\begin{equation}\label{EFElow}
    R_{\mu\nu} = -\left( R_{\mu\nu}^{[2]}\right)^{\text{Low}} +  \kappa^2 \left (T_{\mu\nu} -\frac{1}{2}g_{\mu\nu} T \right)^{\text{Low}},
 \end{equation}
\begin{equation}\label{EFEHigh}
    R_{\mu\nu}^{[1]} = - \left(R_{\mu\nu}^{[2]}\right)^{\text{High}} +  \kappa^2 \left (T_{\mu\nu} -\frac{1}{2}g_{\mu\nu} T \right)^{\text{High}}.
\end{equation}
It is the equation (\ref{EFElow}) that determines the dynamics of the background metric $g{_{\mu\nu}}$ and the energy momentum tensor associated with the high frequency mode $h_{\mu\nu}$. The order of magnitude of the high frequency field equations suggests that $R_{\mu\nu}^{[1]} \sim \partial^{2}h \sim h/\ell^{2}$ while  $R_{\mu\nu}^{[2]} \sim \partial^{2}h^{2} \sim h^{2}/\ell^{2}$. Therefore, $\left( R_{\mu\nu}^{[2]}\right)^{\text{High}}$ is quadratic in the perturbation and negligible with respect to $R_{\mu\nu}^{[1]}$ and thus can be neglected in the linear order equations. Similarly, the analogous estimation applied to the right hand side of \eqref{EFEHigh} provides that $T_{\mu\nu}$ is a smooth macroscopic external matter and its high frequency contribution can arise only via its dependence on $g_{\mu\nu}$ and via the term $g_{\mu\nu}T$. Since it has the order $\mathcal{O}(h)$,
\begin{equation}\label{EMTHigh}
    \left (T_{\mu\nu} -\frac{1}{2}g_{\mu\nu} T \right)^{\text{High}} \sim \mathcal{O}(\frac{h}{L_{B}^{2}}).
\end{equation}
In comparison to $R_{\mu\nu}^{[1]} \sim h/\ell^{2}$  we see that (\ref{EMTHigh}) is smaller than $R_{\mu\nu}^{[1]}$ by a factor of $\mathcal{O}(\ell^{2}/L_{B}^{2})$.
In summary, the low frequency part of the Einstein equations describes the effect of external matter and GWs which affect the curvature of the background, while the high frequency part reduces to the wave equation in this background
\begin{equation}\label{waveeq}
     \Box h_{\mu\nu} -2h_{\alpha\beta} R^{\alpha} {_{\mu\nu}} ^{\beta} = 0.
\end{equation} 
where $\Box=\nabla_\mu \nabla^{\mu}$ as well as $R^{\alpha} {_{\mu\nu}} ^{\beta}$ are computed from the Levi-Civita connection of the background metric $g_{\mu\nu}$.
The above equation can be studied by the \textit{Wentzel-Kramers-Brillouin} (WKB) approximation\footnote{Also known as the eikonal approximation or stationary phase approximation in the literature.}.

The wave equation \eqref{waveeq} cannot be solved explicitly as the approximation of plane waves is not valid in curved spacetime. However, it is to be noted that in the geometric optics limit the plane wave approximation is still valid \cite{Cusin:2019rmt}.
Therefore, the eikonal ansatz can be used to study the effects of background on the propagation of GWs in geometric optics limit and beyond. Then, if the information about the GW amplitude and polarization is carried by the complex symmetric tensor denoted as $\xi_{\mu\nu}$, the high frequency fluctuation can be written in the following form
\begin{equation}\label{Eansatz}
    h_{\mu\nu} =  \mathfrak{Re}\Big\{\Big[\xi^{(0)}_{\mu\nu} +  \epsilon\xi^{(1)}_{\mu\nu}+ \epsilon^{2}\xi^{(2)}_{\mu\nu}+....\Big]e^{i \Phi/\epsilon}\Big \},
\end{equation}
where $\Phi(x)$ is a real scalar function that depends on the coordinates and defines the phase of the GWs while $\epsilon$ is an expansion book-keeping parameter.
The geometric optics limit corresponds to $\epsilon \rightarrow 0 $. As a consequence of this ansatz, independently of the systematic expansion in ${h}_{\mu\nu}$ there exists a derivative expansion of high frequency modes given by the parameter $\epsilon = \ell / L_{B}$.  The hierarchy of the terms are as follows $\epsilon \gg h\epsilon \gg h\epsilon^{2} ... \gg h^{2}$. Therefore, each derivative of the high frequency fluctuation $h_{\mu\nu} $ of (\ref{Eansatz}) will collect a factor $1/ \epsilon$, allowing to separate out geometric optics and beyond geometric optics terms. To study the evolution of the fluctuations up to a required order, one applies the eikonal ansatz (\ref{Eansatz}) in the wave equation \eqref{waveeq} 
\begin{align}\label{WKBexpan}
   &
  e^{i \Phi/ \epsilon}\Big\{ \frac{1}{\epsilon^{2}}[-k^{\beta}k_{\beta}\xi^{(0)}_{\mu\nu}]  \\
  &+ \frac{1}{\epsilon}[i(\nabla_{\beta}k^{\beta}\xi^{(0)}_{\mu\nu} + k^{\beta}\nabla_{\beta}\xi^{(0)}_{\mu\nu}) -k^{\beta}k_{\beta}\xi^{(1)}_{\mu\nu}] +\mathcal{O}(\epsilon^{0}) \nonumber
   \Big\}  = 0,
\end{align}
where we have defined the wave vector $k^\mu$ as the gradient of $\Phi$, that is, $k^{\mu} = g^{\mu\nu} \partial_{\nu}\Phi$. Applying the ansatz to the Hilbert gauge condition $\nabla^{\mu}h_{\mu\nu} = 0$ yields
\begin{align}\label{gaugeEik}
     e^{i \Phi/\epsilon}\Big\{ \frac{1}{\epsilon}[ik^{\mu} \xi^{(0)}_{\mu\nu} ] + \mathcal{O}(\epsilon^{0}) \Big\} =0\,.
\end{align}
Considering the leading $\mathcal{O}(\epsilon^{-1})$ term in \eqref{gaugeEik}, and the leading order $\mathcal{O}(\epsilon^{-2})$ term with the next to the leading order $\mathcal{O}(\epsilon^{-1})$ term in \eqref{WKBexpan} defines the geometric optics limit. Beyond the optical limit terms $\mathcal{O}(\epsilon^{0})$ were considered in \cite{Dalang:2021qhu,Cusin:2019rmt}.

Therefore, we see from \eqref{gaugeEik} that at the leading order, the GW polarization is transverse to the direction of propagation in the geometric optics limit,
\begin{equation}
    k^{\mu} \xi^{(0)}_{\mu\nu} = 0.
\end{equation}
Similarly, the leading order term $ \mathcal{O}(\epsilon^{-2}) $ in the expansion  \eqref{WKBexpan} provides us with the information that the wave vector is a null vector field, $ k_{\beta}k^{\beta} = 0$, hence the gravitational waves travel at the speed of light. Taking a covariant derivative of this relation and making use of the fact that the wave vector is a gradient of the phase function $\Phi$ gives us the geodesic equation 
\begin{equation}\label{EP}
    k^{\mu}\nabla_{\mu}k_{\nu} = 0,
\end{equation}
which tells us that the gravitational wave vector is parallel propagated along the null geodesic of the background metric $g_{\mu\nu}$. It is easy to see that by defining $k^{\mu} = dx^{\mu}/d\lambda$, where $\lambda$ is an affine parameter along the geodesic, and substituting it in the equation (\ref{EP}) we get the geodesic equation in the familiar form
 \begin{equation}\label{Egeo}
    \frac{d^{2}x^{\beta}}{d\lambda^{2}} + \Gamma^{\beta}_{\alpha\mu}\frac{dx^{\alpha}}{d\lambda}\frac{dx^{\mu}}{d\lambda} = 0.
\end{equation}
where $\Gamma^{\beta}_{\alpha\mu}$ are the coefficients of the  Levi-Civita connection of the background metric $g_{\mu\nu}$.
The next-to-leading order provides the following relation
\begin{equation}\label{Epol1} 
    2k^{\beta}\nabla_{\beta}\xi^{(0)}_{\mu\nu} + \nabla_{\beta}k^{\beta}\xi^{(0)}_{\mu\nu} = 0
\end{equation}
In order to understand this equation better, let us rewrite the tensor $\xi^{(0)}_{\mu\nu}$ as  
\begin{equation}
\xi^{(0)}_{\mu\nu} = \mathcal{A} \mathcal{A}_{\mu\nu},
\end{equation} 
where $\mathcal{A}$ is the amplitude, defined as $\mathcal{A} = \sqrt{\xi^{*}_{\mu\nu}\xi^{\mu\nu}}$ where $\xi^{*}_{\mu\nu}$ is the complex conjugate of $\xi^{\mu\nu}$ , and $\mathcal{A}_{\mu\nu}$ is the normalized polarization tensor. Using this decomposition and the gauge condition in the equation (\ref{Epol1}), we obtain the following relations for the amplitude and the  polarization tensor:
\begin{align}
\label{AmpEvol1}\nabla_{\rho}(k^{\rho}\mathcal{A}^{2})  &=  0,\\
\label{parallelT1} k^{\alpha}\nabla_{\alpha}\mathcal{A}_{\mu\nu} &= 0.
\end{align}
The equation \eqref{parallelT1} suggests that in the geometric optics limit the polarization tensor is parallel transported along the gravitational wave direction $k^{\mu}$. Moreover, the equation \eqref{AmpEvol1} can be written as a conservation equation. Let us define the momentum of the gravitons as $P^{\mu} = \hbar k^{\mu}$. Then, the graviton number density along the geodesic bundle is given by $N^{\mu} = \frac{\mathcal{A}^{2}}{\hbar^{2}}P^{\mu}$. Applying it into \eqref{AmpEvol1} leads to the conservation of graviton number density 
\begin{equation}
    \nabla_{\mu}N^{\mu} = 0
\end{equation}
in the geometric optics limit \cite{Gravitational_lenses1992,Maggiore:2007ulw} .
This arises as a consequence of the absence of high frequency fields that contribute to $\left(T_{\mu\nu}\right)^{\text{High}}$ in the Einstein field equations \cite{Garoffolo:2019mna}.

\subsection{Palatini \texorpdfstring{$f(\hat{R})$ gravity}{Lg}}

To obtain all basic relations which govern the propagation of gravitational waves in Palatini $f(\hat{R})$ gravity, we need to perturb the field equations which we obtained in the section \ref{pal}. It is more convenient to start from the conformal representation 
%of Palatini $f(\hat{R})$ in the Einstein frame 
and use the perturbed form of the metric  
\begin{equation}\label{Ppert}
\hat{g}{}^{\text{total}}_{\mu\nu} = \hat{g}_{\mu\nu} +  \hat{h}_{\mu\nu} 
\end{equation}
where $\hat{g}{}_{\mu\nu} =  f'g{}_{\mu\nu}$ and ${\hat{h}}_{\mu\nu} =  f'h_{\mu\nu}  + f''g_{\mu\nu}\delta \hat{R}$, cf.\ \eqref{confmetric} and \eqref{g and h}. The inverse metric is defined by $\hat{g}{}^{\mu\nu}_{\text{total}} = \hat{g}^{\mu\nu} - \hat{h}^{\mu\nu}$. As already mentioned at the end of the section \ref{pal}, since we are dealing with no anisotropic stress, the perturbations coincide and the second term can be neglected, that is, $ \hat{h}_{\mu\nu}= {h}_{\mu\nu}$ \cite{jimenez2015tensor,jimenez2017gravitational,jimenez2018born}.

Substituting the perturbation \eqref{Ppert} in the equation (\ref{P_scalartensor}), neglecting the ``potential'' (effective cosmological constant) contribution and considering only the high frequency part of the Palatini $f(\hat{R})$ gravity yields
\begin{equation}\label{GWmaster1}
   \hat{\Box}{} h_{\mu\nu} - 2{h}_{\alpha\beta}\hat{R}^{\alpha}{}_{\mu\nu}{}^{\beta}  = 0 \,,
\end{equation}
where $\hat{\Box}$ and $\hat{R}^{\alpha}{}_{\mu\nu}{}^{\beta}$ are computed from the Levi-Civita connection $\hat{\Gamma}$ of the long wavelength background metric $\hat{g}$.
Notice that, in this case, the absence of matter
does not reduce Palatini theory to GR. The slowly varying part of the field equations does not have a vanishing trace of the energy-momentum tensor as it happens in the GR. As before, applying the eikonal ansatz one obtains the following %eikonal 
expansion similar to the GR one:  
\begin{equation}
\begin{split}
    e^{i \Phi/ \epsilon}\Big\{ \frac{1}{\epsilon^{2}}[-\hat{k}^{\beta}\hat{k}_{\beta}\xi^{(0)}_{\mu\nu}] +
    \frac{1}{\epsilon}[i(\hat{\nabla}_{\beta}\hat{k}^{\beta}\xi^{(0)}_{\mu\nu}\\ + \hat{k}^{\beta}\hat{\nabla}_{\beta}\xi^{(0)}_{\mu\nu})-\hat{k}^{\beta}\hat{k}_{\beta}\xi^{(1)}_{\mu\nu}] + \mathcal{O}(\epsilon^{0}) \Big \} = 0
    \,,
\end{split}
\end{equation}
where  the covariant derivatives $\hat{\nabla}$ are defined with respect to the connection $\hat{\Gamma}$. 
We see that the leading order provides that the GWs in Palatini $f(\hat{R})$ gravity also follow null paths defined by the wave vector $\hat{k}_{\mu}\hat{k}^{\mu}=0$. On the other hand, the geodesic equation requires a bit of discussion. 
Taking the covariant derivative (with respect to the connection $\hat{\Gamma}$) of this leading order term gives an autoparallel
\begin{equation}
       \hat{k}^{\mu}\hat{\nabla}{_{\mu}}\hat{k}_{\nu} = 0 \,,
\end{equation}
which can be rewritten in a familiar form 
\begin{equation}\label{Pgeo}
    \frac{dx^{\beta}}{d\hat{\lambda}^{2}} + \hat{\Gamma}^{\beta}_{\alpha\mu}\frac{dx^{\alpha}}{d\hat{\lambda}}\frac{dx^{\mu}}{d\hat{\lambda}} = 0
\end{equation}
by defining the wave vector as $\hat{k}^{\mu} = dx^{\mu}/d\hat{\lambda}$, where $\hat{\lambda}$ is the affine parameter in the conformal frame. This is a geodesic equation of the conformal metric $\hat{g}$ which we introduced as a convenient calculation tool in the theory. In principle the geodesics of $\hat{g}$ can be contrasted with the geodesics of the original metric $g$, and it is the latter that is obeyed by the matter particles (since the matter action \eqref{action} depends on the metric $g$). However, due to the conformal invariance of the null directions, the null geodesics of $g$ and $\hat{g}$ coincide, and thus in the geometric optics limit the GWs in GR and Palatini $f(\hat{R}$) gravity follow the same paths.\footnote{Consider the conformal relations for the wave vectors $\hat k^\mu= (f'(\hat R))^{-1} k^\mu$ and  $\hat k_\mu=k_\mu$ \cite{Stachowski:2016zio,fleurry_thesis}.}

Although the next to leading order $\mathcal{O}(\epsilon^{-1})$ also seems to be similar,
\begin{equation}\label{P_amp_evol}
    2k^{\alpha}\hat{\nabla}_{\alpha}\xi^{(0)}_{\mu\nu} + \hat{\nabla}_{\alpha}\hat{k}^{\alpha}\xi^{(0)}_{\mu\nu} = 0,
\end{equation}
let us decompose the wave tensor $\xi^{(0)}_{\mu\nu}$ as previously in order to understand the physical consequences. The polarization tensor $\mathcal{A}_{\mu\nu}$ in this theory is parallel propagated along the %propagation 
direction of $\hat{k}^{\alpha}$,
\begin{equation}
    \hat{k}^{\alpha}\hat{\nabla}_{\alpha}\mathcal{A}_{\mu\nu} = 0.
\end{equation}
{However, since due to the conformal relationship the vectors $\hat{k}^\alpha$ and $k^\alpha$ are parallel, the polarization tensor is also parallel propagated along the light cone direction $k^\alpha$ of the original metric $g$, like the light rays.}

Moreover, we also obtain information about the conservation of graviton number density. It goes as follows: if we define the momentum of gravitons as $\hat{P} = \hbar \hat{k}^{\mu}$, then graviton number density along a geodesic ray bundle can be defined as $\hat{N}^{\mu} = \frac{\mathcal{A}^{2}}{\hbar^{2}}\hat{P}^{\mu}$. Therefore, as we have seen in the case GR (\ref{AmpEvol1}), one can obtain a relation for the amplitude as 
\begin{equation}\label{Pal_evol}
    2\hat{k}^{\alpha}\hat{\nabla}_{\alpha}\mathcal{A} + \mathcal{A} \hat{\nabla}_{\alpha}\hat{k}^{\alpha} = 0 \,.
\end{equation}
Rewriting the above equation as $\hat{\nabla}_{\mu}(\hat{k}^{\mu}\mathcal{A}^{2})= 0$ and making use of the expression for graviton number density we obtain \begin{align}
    \hat{\nabla}_{\mu}\hat{N}^{\mu} = 0 \,.
\end{align}
Note that the graviton number density conservation in  Palatini $f(\hat{R})$ gravity holds for the quantity $\hat{N}$ and conformal metric $\hat{g}$  (in terms of its Levi-Civita covariant derivative $\hat\nabla_\mu$). If we rewrite these formulas for the quantity ${N}^{\mu}$ and the vector $k^{\mu}$ as in GR, the conservation of graviton number density turns out to be violated. Namely, one has
\begin{equation}
    2{k}^{\alpha}{\nabla}_{\alpha}\mathcal{A} + \mathcal{A} {\nabla}_{\alpha}{k}^{\alpha} +\mathcal{A} k^\alpha\partial_\alpha \ln f'(\hat R)= 0.
\end{equation}
which can be rewritten as
\begin{equation}\label{cons_Pal}
    \nabla_{\mu}N^{\mu} =  - N^{\mu} \partial_{\mu} \ln f'.
\end{equation}
This feature can be compared to the conservation of matter energy-momentum in terms of $\nabla_\mu$, \eqref{eq: matter conservation}, but not in terms of the conformal $\hat{\nabla}_\mu$, \eqref{eq: matter nonconservation}. For gravitational waves the roles of $\nabla_\mu$ and $\hat{\nabla}_\mu$ have switched, which is one of the key features of Palatini $f(\hat{R})$ gravity \cite{jimenez2015tensor,jimenez2018born}. Later in Sec.\ \ref{subsec: GW amplitude evolutionm} we see that the graviton number nonconservation can be attributed to the change in amplitude. Such effects are also observed in scalar-tensor theories \cite{Garoffolo:2019mna} due to the presence of high frequency scalar fields. Further, the non-conservation has consequences related to the observable such as GW luminosity distance \cite{Deffayet:2007kf, Saltas:2014dha, Nishizawa:2017nef, Belgacem:2017ihm}.

\section{Gravitational wave amplitude evolution} \label{GWampl}

Passing by distributions of matter can affect the amplitude and polarisation of a gravitational wave, as well as deflect the direction of the rays. We will focus on the amplitude first.

\subsection{Newman-Penrose tetrad in both theories}\label{NP_newbasis}

The analysis of the polarization states and the evolution of amplitude can be made more simple by introducing a Newman-Penrose (NP) tetrad \cite{Newman:1961qr, Chandrasekhar:1985kt} of null co-vectors
\begin{equation}\label{NP tetrad}
    e^{A}_{\mu}=  \{ k_{\mu},m_{\mu},l_{\mu},n_{\mu} \}\,,
\end{equation}
which are parallel transported along the null geodesics. The vectors $k^{\mu}$ and $n^{\mu}$ are real while $m^{\mu}$ and $l^{\mu}$ are complex conjugate pairs whose only non-vanishing contractions satisfy the relations:
\begin{equation}
    m_{\mu}l^{\mu} = 1, \hspace{0.5cm}  k_{\mu}n^{\mu}  = -1.
\end{equation}

It can be shown that they can be used to express the spacetime metric as
\begin{equation}
    g_{\mu\nu} =  m_{\mu}l_{\nu} + m_{\nu}l_{\mu}  - n_{\mu}k_{\nu} - n_{\nu}k_{\mu}.
\end{equation}
The GW is represented by a second-rank symmetric tensor. Therefore, ten independent components of the tensor can be rewritten in terms of a symmetric basis constructed from the null tetrad \cite{Cusin:2019rmt} 
\begin{equation}
    \Theta^{AB}_{\mu\nu} =  \frac{1}{2}(A_{\mu}B_{\nu} + A_{\nu}B_{\mu}),
\end{equation}
where $A_\mu,B_\mu$ have the same structure as $e_{\mu}$ in \eqref{NP tetrad}.
%$A_\mu,B_\mu = \{ k_{\mu},m_{\mu},l_{\mu},n_{\mu} \}$.
The tensor $\xi_{\mu\nu}$ can be then expanded in this basis as follows
\begin{equation}
\begin{split}
   & \xi_{\mu\nu} = \mathcal{C}_{kk}\Theta^{kk}_{\mu\nu} +  \mathcal{C}_{ll}\Theta^{ll}_{\mu\nu} +  \mathcal{C}_{mm}\Theta^{mm}_{\mu\nu} +  \mathcal{C}_{nn}\Theta^{nn}_{\mu\nu} +  \mathcal{C}_{kl}\Theta^{kl}_{\mu\nu}\\ & \quad +  \mathcal{C}_{km}\Theta^{km}_{\mu\nu} +  \mathcal{C}_{kn}\Theta^{kn}_{\mu\nu} +  \mathcal{C}_{ml}\Theta^{ml}_{\mu\nu} +  \mathcal{C}_{nl}\Theta^{ml}_{\mu\nu} +  \mathcal{C}_{mn}\Theta^{mn}_{\mu\nu},
    \end{split}
\end{equation}
where $\mathcal{C}_{AB}$ are complex coefficients. In the geometric optics limit, after using all gauge freedom this tensor reduces to the following form 
\begin{equation}\label{np_polarisation}
    \xi^{(0)}_{\mu\nu} = \mathcal{C}^{(0)}_{ll}l_{\mu}l_{\nu} +\mathcal{C}^{(0)}_{mm}m_{\mu}m_{\nu},
\end{equation}
which reveals the two polarizations. To see this let us substitute $l^{\mu} = (x^{\mu} + i y^{\mu})/\sqrt{2}$ and $m^{\mu} = (x^{\mu} - i y^{\mu})/\sqrt{2}$, whereby \eqref{np_polarisation} can be rewritten as 
\begin{equation}\label{wavetensor}
    \xi^{(0)}_{\mu\nu}  =  H_{+}\mathcal{A}^{+}_{\mu\nu} + H_{\times}\mathcal{A}^{\times}_{\mu\nu} \,.
\end{equation}
Here $\mathcal{A}^{+}_{\mu\nu} = x_{\mu}x_{\nu}-y_{\mu}y_{\nu}$ and $\mathcal{A}^{\times}_{\mu\nu} = x_{\mu}y_{\nu} + y_{\nu}y_{\nu}$
are the two polarization tensors in GR and $H_{+}, H_{\times}$ are their corresponding amplitudes. The complex amplitudes $\mathcal{C}_{AB}$ are related to the polarizations amplitudes  as follows
\begin{equation}
    H_{+} = \frac{1}{2}( \mathcal{C}^{(0)}_{ll}+\mathcal{C}^{(0)}_{mm})  \; ; \; H_{\times} =  \frac{i}{2}( \mathcal{C}^{(0)}_{ll}-\mathcal{C}^{(0)}_{mm}).
\end{equation}

\subsection{General relativity} \label{In GR}

Gravitational waves in the geometrical optics limit are rays of null congruences. The congruence is a set of parametrized curves such that only one curve passes through each point. The behavior of one null geodesic with respect to another reference ray is governed by the separation vector $\chi^{\mu}$. The evolution of $\chi^{\mu}$ depends on the background Riemann tensor $R_{\mu\nu\rho\sigma}$ %\re{$R^{\sB}_{\mu\nu\rho\sigma}$?} 
and obeys the geodesic deviation equation 
\begin{equation}\label{deviation}
    \frac{D^{2}\chi^{\mu}}{d\lambda^{2}} = R^{\mu}{}_{\nu\rho\sigma}k^{\nu}k^{\rho}\chi^{\sigma},
\end{equation}
where $\lambda$ denotes the affine parameter and $D$ is the directional covariant derivative, that is, $k^\mu\nabla_\mu$ with respect to the the null geodesics $k^\mu$. It can be written in the matrix form (see Appendix \eqref{appendix_B}) and it is also known as the Sachs equation when projected on a two-dimensional spacelike surface (see the discussion below).

The cross-section of the beam has a nonzero extension except from the point of observation where the rays converge and the separation vector $\chi(\lambda_{O}) = 0$, with $\lambda_{O}$ being the value of the affine parameter at the observation point. Therefore, a given observer with four-velocity $u^{\mu}$ can also project the separation vector into a 2-dimensional surface which is called a screen space. The screen space is perpendicular to the line of the sight direction $d^{\mu}$ of the observer, with $d^{\mu}d_{\mu}=1$ and $d^{\mu}u_{\mu}=0$. The screen projector is defined as 
\begin{equation}
    Q^{\mu\nu} =  g^{\mu\nu} + u^{\mu}u^{\nu} - d^{\mu}d^{\nu} \,.
\end{equation}

To study the morphology of the beam in the screen space by an observer with $u^{\mu}$, let us consider the physical area $A$ of the cross-sections of the beam on the screen space. It is given by the following expression
\begin{equation}\label{cross}
A  =  \int d \chi^{1} d \chi^{2} = \int \text{det} \mathcal{D} d\dot{\chi}^{1}_{O} d\dot{\chi}^{2}_{O},     
\end{equation}
where $\mathcal{D}$ is the Jacobi matrix satisfying the equation \eqref{jacobi_mat_eqn}, where $\dot{\chi}^{1}_{O}$ denotes the derivative with respect to the affine parameter.

Let us now consider an infinitesimal light beam. Then, the Jacobi matrix $\mathcal{D}$ can be regarded as a constant, and the evolution rate of the area of the cross-sections of the beam \eqref{cross} with respect to the affine parameter $\lambda$ can be expressed with the use of the expansion scalar $\theta$ (see its definition in the Appendix \ref{appendix_B})
$$ 
\frac{\dot{A}}{A} =  \frac{1}{\det\mathcal{D}} \frac{d (\det \mathcal{D})}{d\lambda} =  \text{tr}(\mathcal{S})=2\theta.
$$ 

Therefore, the expansion scalar $\theta$ can be interpreted as the evolution rate of the beam’s area:
\begin{equation}\label{expD}
    \theta =  \frac{1}{D_{A}}\frac{d D_{A}}{d\lambda},
\end{equation}
where the angular diameter distance $D_A$ is proportional to $\sqrt{A}$ \cite{Fluerry_2015}. Now on, using the above relations along with amplitude evolution  equation \eqref{Epol1} rewritten in terms of two polarization states in GR \eqref{np_polarisation}, we get
\begin{equation}
    2k^{\mu}\nabla_{\mu}\mathcal{C}_{AB} + \nabla_{\mu}k^{\mu}\mathcal{C}_{AB} = 0
\end{equation}
and
\begin{equation}
    \frac{d \ln \mathcal{C}_{AB} }{d\lambda} + \frac{d \ln D_{A}}{d \lambda} = 0 \,.
\end{equation}
Integrating this equation from the source to the observation point we get the amplitude of polarization at an arbitrary position in terms of the affine parameter $\lambda$ 
\begin{equation}\label{GR_amp_evolution}
    \mathcal{C}^\text{GR}_{AB}(\lambda) = \frac{\mathcal{C}_{AB}(\lambda_{s})D(\lambda_{s})}{D(\lambda)} \,,
\end{equation}
where $\lambda_{s}$ is the affine parameter at the source position.

\subsection{Palatini \texorpdfstring{$f(\hat{R})$ gravity}{Lg}}

As mentioned in earlier sections of Palatini $f(\hat{R})$ gravity, the metric $\hat{g}$ that is convenient to describe the gravitational dynamics in this theory is related to the physical metric $g$ via the conformal factor $f'(\hat{R})$. Therefore all quantities that describe the morphology of the beam, including the Sachs vector equations and expansion, carry an imprint of this conformal factor. The screen projector is given by $\hat{Q}_{\mu\nu} =  f'  Q_{\mu\nu} $,
this is due to the re-scaling of the line of sight direction and the four-velocity of the observer\footnote{Notice that $\hat{u}$ is not a physical observer - the observer vector field is related to the metric $g$, and in our case, it co-moves with the fluid given by the energy-momentum tensor $T_{\mu\nu}$.}
\begin{equation}
    \hat{d}_{\mu} =   \sqrt{ f'} d_{\mu} \,, \qquad \hat{u}_{\nu} =  \sqrt{ f'} u_{\nu} \,.
\end{equation}
As a consequence of this conformal transformation, the expansion term \eqref{expD} gets an extra term which is related to the conformal factor and hence to the trace of the energy-momentum tensor via the trace equation \eqref{trace}. Therefore the Ricci and Weyl lensing terms get modified in Palatini $f(\hat{R})$ gravity. The expansion scalar in Palatini $f(\hat{R})$ gravity is related to the expansion scalar in general relativity as 
\begin{equation}\label{theta_hat}
    \hat{\theta} = \frac{1}{2}\hat{\nabla}_\alpha\hat{k}^\alpha= \frac{1}{2}\frac{d \ln f' }{d\lambda} + \frac{1}{f'}\theta
\end{equation}
where $\theta$ is related to the angular diameter distance \eqref{expD}. 

Using the relation for the evolution of amplitude in Palatini $f(\hat{R})$ gravity given by \eqref{P_amp_evol} and making use of the new basis introduced in the section (\ref{NP_newbasis}) and  the gauge freedom, we obtain the following equation for the evolution of amplitude
\begin{equation}
     2\hat{k}^{\mu}\hat{\nabla}_{\mu}\mathcal{C}_{AB} + \hat{\nabla}_{\mu}\hat{k}^{\mu}\mathcal{C}_{AB} = 0
\end{equation}
where $\mathcal{C}_{AB}$ denotes the amplitude of two polarizations. It is to be remembered that the wave direction is given by $\hat{k}^{\mu} = \frac{k^{\mu}}{f'}$ and the covariant derivative $\hat\nabla$ is given by the connection $\hat\Gamma$ in Palatini $f(\hat{R})$ gravity. Integrating the above equation from the source to the observation point and making use of \eqref{theta_hat} one gets the evolution of the gravitational wave amplitude in this theory of gravity,
\begin{align}\label{P_amp_evolution}
   & \mathcal{C}^\text{Palatini}_{AB} \\
    &= \mathcal{C}^\text{GR}_{AB}(\lambda) \exp \Big[ -\int_{\lambda_{s}}^{\lambda_{o}}\Big(2 \frac{d \ln f'}{d \lambda}+\frac{1}{2} f' \frac{d \ln f'}{d\lambda} \Big)d\lambda \Big], \nonumber
\end{align}
where $ \mathcal{C}^\text{GR}_{AB}(\lambda)$ is given by \eqref{GR_amp_evolution} where $\lambda_{o}$ and $\lambda_{s}$ are the value of the affine parameter at the observer and source respectively.
This is our main result. We observe that in the geometric optics limit the two polarizations evolve independently, in the same way as in GR. However, in Palatini $f(\hat R)$ gravity we are dealing with the additional term making the amplitude decrease with the distance. The exponential term is related to the mass distribution of the lens through the conformal factor. This extra exponential factor modifies the amplitude of GW detected, hence modifying the luminosity distance inferred from the signal based on the GR prediction. A similar effect is also seen in scalar-tensor theories \cite{Garoffolo:2019mna, Tasinato:2021wol} where this modification can be interpreted as a correction to the GW luminosity distance. Alternatively, it is also possible to derive the luminosity distance starting from the energy-momentum tensor similar to \cite{Sasaki:1987ad,Tasinato:2021wol}(see Appendix \eqref{luminosity_distance}). The corrected luminosity distance in Palatini theory takes the following form 
\begin{equation}\label{luminosity_distance_corr}
    d^{\text{P}}_{L} = d^{\text{GR}}_{L}\exp \Big[ \int_{\lambda_{s}}^{\lambda_{o}}\Big(2 \frac{d \ln f'}{d \lambda}+\frac{1}{2} f' \frac{d \ln f'}{d\lambda} \Big)d\lambda \Big]
\end{equation}

As mentioned in Sec.\ \ref{go}, this is a consequence of the non-conservation of graviton number density. Therefore in the event of GW lensing in Palatini $f(\hat{R})$ one observes a larger luminosity. Further, in the case of a multi-messenger GW event with the electromagnetic follow-up  observation, 
luminosity distance measured in the electromagnetic window would be different from that measured using GW standard sirens.

\subsection{Nonrelativistic limit of Palatini \texorpdfstring{ $f(\hat{R})$ gravity}{Lg}}

In what follows, we will consider a non-relativistic object described by the Palatini $f(\hat R)$ gravity, such as a star or galaxy. To do so, we need to briefly discuss the non-relativistic limit of the theory. For an analytic function $f(\hat {R})=\sum_{i=0}\alpha_i\hat{R}^i$, the Poisson equation is given by \cite{Toniato:2019rrd,Hernandez-Arboleda:2023abv}
\begin{equation}\label{poisson}
    \nabla^2\Phi\approx 4\pi G(\rho +2\alpha\nabla^2\rho),
\end{equation}
where $\alpha$ comes from the quadratic term of the Lagrangian (we neglect cosmological constant). Therefore, in the non-relativistic objects, the only correction to the GR equations which has a negligible effect is given by the polynomial term
\begin{equation}
\label{eq: Starobinsky Palatini}
    f(\hat R)= \hat R + \alpha \hat R^2.
\end{equation}
Inserting this functional into the trace equation \eqref{structural} provides the analogous to GR relation between the trace of the energy-momentum tensor and the Palatini curvature scalar:
\begin{equation}
\label{eq: Starobinsky Palatini R T}
    \hat R = - \frac{8\pi G}{c^4} T,
\end{equation}
where $T=3p - c^2\rho$  if we consider a perfect-fluid description of matter. Since we are interested in  non-relativistic objects, pressure, when it appears together with the energy density, is negligible in such objects, that is, $p/c^2 \ll \rho$. Therefore, any modification introduced by Palatini gravity in the Poisson equation \eqref{poisson} and further in the gravitational potential, is signalized by the functions of the energy density $\rho(r)$ (with the parameter $\alpha$).

\subsection{Gravitational wave amplitude evolution}\label{subsec: GW amplitude evolutionm}

The above set of equations can be used to evaluate gravitational wave amplitude evolution and hence the luminosity distance correction in Palatini $f(\hat{R})$ gravity. From Eqs.\ \eqref{eq: Starobinsky Palatini}, \eqref{eq: Starobinsky Palatini R T} the conformal factor  becomes $f' =  1 +  2\alpha \hat{R} = 1 + A\rho$, where $A= 2 \alpha \kappa^2 c^{2}$. As a first approximation, one can rewrite the amplitude evolution expression \eqref{P_amp_evolution} in the form 
\begin{equation}
    \mathcal{C}^{P}_{AB} =  \mathcal{C}^{GR}_{AB}\exp \Big[ - \frac{5A}{2}\int_{\rho_{s}}^{\rho_{o}} d \rho \Big], 
\end{equation}
 It is quite clear from \eqref{luminosity_distance_corr} that the above equation can be expressed as the correction to luminosity distance. 
Any deviation from GR comes from the exponential factor present in the above equation. 

In order to see the effect of Palatini gravity on the amplitude evolution, we need to consider some density profile which describes our lens. As Palatini $f(\hat{R})$ gravity differs from GR only in the presence of matter, it is evident from above that the most popular lens model considered in the literature, that is, a point like mass in the vacuum \cite{Gravitational_lenses1992}, will not give any additional effect compared to GR. Because of that fact, we need to consider an extended body with a given density profile that resides asymmetrically on the path of the gravitational ray, whereby the entry and exit values of the density differ. For instance, we may consider a GW source close to the center  of a galaxy cluster and estimate the accumulated extra effect as the wave exits the cluster and passes into the cosmic intergalactic medium.   
Computing the terms in the exponential for Navarro-Frenk-White(NFW) density profile \cite{Navarro:1996gj} for an XCOP galaxy cluster sample \cite{Ettori:2018tus} A85 with a radial distance of $b = 3 \text{ Mpc}$ from the cluster center one obtains the magnitude of the correction factor to be of the order $10^{-42}$ for $\alpha \sim 10^{9}$ $\mathrm{m^{2}}$. Analogously, assuming a SIS  for the same radial distance one estimates the correction to be of the order $10^{-43}$.  Therefore it can be seen that the correction to the GW amplitude as well as to the luminosity distance during the wave propagation is negligible and clearly beyond the sensitivity limit of the GW detectors. Out of curiosity, we may also estimate the accumulated extra effect for an ancient wave that has been travelling  since the early universe. Integrating over the cosmic mean density from a redshift $z_{s} = 3000$ till now will have the luminosity distance corrected by a factor of $10^{-33}$, which is still negligible.  However, if one uses the upper constraint for $\alpha \sim 10^{49} \ \mathrm{m^{2}}$ from SNIa and BAO \cite{Gomes:2023xzk} the corrections for the GW luminosity distance is $10^{-2}$ which is non-negligible.

\section{Gravitational wave deflection angle and rotation of polarization plane}
\label{sectionV}

So far we have discussed the propagation of GW through an extended distribution of matter characterized by some energy density $\rho$ distribution profile. In this section, we discuss how GW propagation is influenced by a compact matter distribution, i.e.\ gravitational lensing.

\subsection{Gravitational lensing}

The subject of gravitational lensing has become a mature field of astrophysics \cite{Gravitational_lenses1992, Meneghetti} having its own rich mathematical formalism \cite{Petters} and numerous practical applications \cite{Treu}. Referring the reader to the literature just cited for a detailed and systematic introduction, we give here a short, concise review of basic concepts of strong lensing and notation used in the sections below. Strong gravitational lensing occurs whenever the source is located at a distance
\footnote{In cosmological scales these distances have a meaning of angular diameter distances.} 
$D_S$, a deflecting mass -- the lens at a distance of $D_L$ from the observer (distance between the source and the lens being $D_{LS}$) are almost perfectly aligned. Then instead of a single image of the source, multiple images can be formed. This is possible because the weak perturbation of the background metric by the lens endows the space with the effective refractive index $n=1 - \frac{2 \Phi}{c^2}$, where $\Phi$ is the Newtonian potential of the lens. 

In practice, the size of the lens is negligible in comparison to the characteristic distances of this optical system and it is convenient to use \textit{thin lens approximation} i.e.\ to assume that all changes in propagation of null geodesics from the source to the observer (with respect to propagation through the background space undistorted by the lens) occurs at the locus of the closest encounter with the lens. This suggests a natural choice of (local) coordinate system: the line of sight to the center of the lens $z$ (optical axis of the system), the lens plane - perpendicular to $z$ at lens location, and the source plane - perpendicular to $z$ at the source location. The intersection of $z$ axis with lens and source planes define the origin of coordinate systems on these planes spanned by arbitrarily (but consistently) chosen orthogonal vectors. Radius vectors (in physical [length] units) on the lens and source planes are denoted $\vec{\xi}$ and $\vec{\eta}$, respectively. Because observations on the sky (the celestial sphere) are expressed in terms of angles, it is convenient to describe strong lensing in the following way. The rays from the source at the intrinsic angular position $\vec{\beta}$, i.e.\ lying on the source plane at a distance $\vec{\eta} = \vec{\beta} D_S$ from the optical axis, impact the lens plane at the position $\vec{\xi} = \vec{\theta} D_L$ and reach the observer because they are deflected by the angle $\vec{\gamma}$. In consequence, the observer registers the source's image at angular position $\vec{\theta}$ instead of intrinsic $\vec{\beta}$, which is governed by the lens equation
\begin{equation}
\vec{\beta} = \vec{\theta} - \vec{\gamma}(\vec{\theta}) \,.
\end{equation}

 It is convenient to choose some length scale $\xi_0$ on the lens plane (and associated to it $\eta_0 = \xi_0 D_S /D_L$ on the source plane), and use dimensionless vectors $\vec{x} = \vec{\xi} /\xi_0$, $\vec{y} = \vec{\eta} /\eta_0$. In this case the lens equation reads $\vec{y} = \vec{x} - \vec{\gamma}(\vec{x})$ and the deflection angle is $\vec{\gamma}=\nabla_{\vec{x}}(\Psi(\vec{x}))$. The dimensionless 2D deflection potential is defined as 
\begin{equation}
\Psi(\vec{\xi}) = \frac{D_L D_{LS}}{\xi_0 D_S} \frac{2}{c^2} \int \Phi(\vec{\xi}, z) dz.   
\end{equation}

Let us finally remark, that although the choice of the length scale $\xi_0$ is arbitrary, it is convenient to use for this purpose the so-called Einstein radius on the lens plane $\xi_0 = \theta_E D_L$, where angular Einstein radius $\theta_E$ satisfies $\theta_E = \vec{\gamma}(\theta_E)$. 
It is also useful to introduce the convergence
\begin{equation}
\kappa(\vec{\xi}) = \Sigma(\Vec{\xi}) / \Sigma_{cr}
\end{equation}
where 
\begin{equation}
\Sigma(\vec{\xi}) = \int_{-\infty}^{\infty} \rho(\vec{\xi},z) dz
\end{equation}
is the surface mass density of the lens and 
\begin{equation}
\Sigma_{cr} = \frac{c^2}{4 \pi G} \frac{D_S}{D_L D_{LS}}
\end{equation}
is the so-called critical surface density \cite{Gravitational_lenses1992}. Dimensionless 2D deflection potential and convergence satisfy the 2D Poisson equation:

\begin{equation}
 \nabla^2_{\vec{x}} \Psi(\vec{x}) = 2 \kappa(\vec{x}). 
\end{equation}

\subsection{General Relativity}

To evaluate the corrections in polarization due to lensing, one needs to perturb the tetrad $ k^{\mu} = \bar{k}^{\mu} + \delta k^{\mu} $, where $\delta k^{\mu}$ is the first order correction which satisfies the perturbed geodesic equation \cite{fleurry_thesis}:
\begin{equation}\label{Pert_geod}
    \bar{k}^{\nu}\nabla_{\nu}\delta k^{\mu} = - \delta \Gamma^{\mu}_{\nu\rho}\bar{k}^{\nu}\bar{k}^{\rho} \,.
\end{equation}
The Christoffel symbols are evaluated for the following weak-field metric
\begin{equation}\label{Pmetric1}
    ds^{2} =  -(1+2\Phi )dt^{2} + (1-2\Phi)(dx^{2} + dy^{2}+ dz^{2}),
\end{equation}
where $\Phi$ is the gravitational potential of the lens\footnote{We have already used the fact that gravitational potentials coincide in the case of GR; see Appendix \ref{appA} for the modified gravity case.}.
Solving \eqref{Pert_geod} for the non-vanishing contributions of the Christoffel components arising from the \eqref{Pmetric1}, one obtains
\begin{equation}\label{delta_x}
   \delta x^{\mu} = \frac{1}{2}\Big(\vec{\gamma} \cdot \vec{x}, -(\vec{\gamma} \cdot\vec{x})\vec{k} \Big ) \,,
\end{equation}
\begin{equation}\label{delta_y}
   \delta y^{\mu} = \frac{1}{2}\Big(\vec{\gamma}\cdot \vec{y}, -(\vec{\gamma} \cdot\vec{y})\vec{k} \Big) \,,
\end{equation}
where $\vec{\gamma}=\nabla_{x}(\Phi(x))$ is the deflection angle due to lensing. 
%, $\Psi(x)$ is the lens potential while $\nabla_{\perp}$ is the projected gradient on the plane perpendicular to the GW propagation. 
The vectors $\vec{x},\vec{y}$  are the spatial parts %space part 
of $x^{\mu},\,y^{\nu}$ introduced in \ref{NP_newbasis}, and $\vec{k}$ denotes the direction of propagation of the GW in space. Now, one can use the basis described in \ref{NP_newbasis} to write the wave tensor in a more familiar form. Therefore, the wave tensor \eqref{wavetensor} can be rewritten by changing the wave vector as $x^{\mu} \rightarrow x^{\mu} + \delta x^{\mu}$,  $y^{\mu} \rightarrow y^{\mu} + \delta y^{\mu}$ arsing from the presence of the lens, where $\delta x$ and $\delta y$ are given by \eqref{delta_x} and \eqref{delta_y}, respectively. The polarization tensor undergoes a rotation and the wave tensor becomes
\begin{equation}
    \xi^{(0)}_{\mu\nu}  =  H_{+}\tilde{\mathcal{A}}^{+}_{\mu\nu} + H_{\times}\tilde{\mathcal{A}}^{\times}_{\mu\nu},
\end{equation}
where $\tilde{\mathcal{A}}^{+}$ and $\tilde{\mathcal{A}}^{\times}$ are the new modified polarization tensors due to the presence of the lens. They are defined as 
\begin{equation}\label{rotated_plus}
\tilde{\mathcal{A}}^{+}_{ij} = \mathcal{A}^{+}_{ij}    -\frac{1}{2}(\vec{\gamma}\cdot\vec{x})\mathcal{A}^{x}_{ij} +\frac{1}{2}(\vec{\gamma}\cdot\vec{y})\mathcal{A}^{y}_{ij}
\end{equation}
and
\begin{equation}\label{rotated_cross}
\tilde{\mathcal{A}}^{\times}_{ij} = \mathcal{A}^{\times}_{ij}    -\frac{1}{2}(\vec{\gamma}\cdot\vec{y})\mathcal{A}^{x}_{ij} -\frac{1}{2}(\vec{\gamma}\cdot\vec{x})\mathcal{A}^{y}_{ij}
\end{equation}
where $\mathcal{A}^{x}_{ij} = x_{i}k_{j}+k_{i}x_{j}$ and $\mathcal{A}^{y}_{ij} = y_{i}k_{j}+k_{i}y_{j}$ are the polarization tensors for the vector -x and vector -y polarizations, respectively. One can see that the effect of the lens is that it introduces additional polarizations in GR, which is an illusion (see e.g.\cite{Hou:2019wdg}). %However, 
Namely, one can always choose appropriate coordinates to remove these non-tensorial polarizations. Therefore, we do not deal here with any additional physical effect.

%%%%%%%

\subsection{Palatini $f(\hat R)$ gravity}
Similarly, to evaluate the perturbed quantities we can use the perturbed geodesic equation similar to the perturbation of light rays (\cite{fleurry_thesis}). Substituting the relation $ k^{\mu} = \bar{k}^{\mu} + \delta k^{\mu} $ into the geodesic equation and evaluating it to  first order in  perturbation we obtain the following relation for the tetrad $\delta k^{\mu}$ (and similarly for other tetrads), 
\begin{equation}\label{pge}
    \bar{k}^{\nu}\hat{\nabla}_{\nu}\delta k^{\mu} = - \delta \hat{\Gamma}^{\mu}_{\nu\rho}\bar{k}^{\nu}\bar{k}^{\rho} \,.
\end{equation}
To evaluate the above equation, one must calculate the non-vanishing components of the affine connection corresponding to the background geometry associated with the lensing mass (see Appendix \ref{appA}). Since we are interested in a static case, the conformal factor $f'(\hat R)=f'(T)$, where $T$ is the trace of the energy-momentum tensor, is independent of time, and hence we will deal with the spatial derivatives only in the above expression. Analogously as in GR, we deal with the following solutions of \eqref{pge}:
\begin{equation}\label{per_tetrads1}
   \delta x^{\mu} = \frac{1}{2}\Big(\vec{\gamma}^{P} \cdot \vec{x}, -(\vec{\gamma}^{P} \cdot\vec{x})\vec{k} \Big )
\end{equation}
\begin{equation}\label{per_tetrads2}
   \delta y^{\mu} = \frac{1}{2}\Big(\vec{\gamma}^{P}\cdot \vec{y}, -(\vec{\gamma}^{P} \cdot\vec{y})\vec{k} \Big),
\end{equation}
where
\begin{equation}\label{lensang}
   \vec \gamma^{P} =  \nabla_{\vec{x}}(f'\Psi(\vec{x}))
\end{equation}
is the gravitational deflection due to the lens in this theory, whereas usually $\Psi(x)$ is the lens potential in a given theory of gravity and the deflection angle $\gamma ^{P}$ reduces to GR when $f' = 1$ ($\alpha=0$ for the considered cases). Recall that $\nabla_{\perp}$ is the projected gradient on the plane perpendicular to the GW propagation. Therefore, the wave tensor \eqref{wavetensor} can be rewritten by changing the wave vector as $x^{\mu} \rightarrow x^{\mu} + \delta x^{\mu}$,  $y^{\mu} \rightarrow y^{\mu} + \delta y^{\mu}$. The polarization tensor undergoes a rotation and the wave tensor becomes
\begin{equation}
    \xi^{(0)}_{\mu\nu}  =  H_{+}\tilde{\mathcal{A}}^{+}_{\mu\nu} + H_{\times}\tilde{\mathcal{A}}^{\times}_{\mu\nu}
\end{equation}
where $\tilde{\mathcal{A}}^{+}$ and $\tilde{\mathcal{A}}^{\times}$ are the new modified polarization tensors defined as 
\begin{equation}\label{rotated_plus_P}
\tilde{\mathcal{A}}^{+}_{ij} = \mathcal{A}^{+}_{ij}    -\frac{1}{2}(\vec{\gamma}^{P}\cdot\vec{x})\mathcal{A}^{x}_{ij} +\frac{1}{2}(\vec{\gamma}^{P}\cdot\vec{y})\mathcal{A}^{y}_{ij}
\end{equation}
and
\begin{equation}\label{rotated_cross_P}
\tilde{\mathcal{A}}^{\times}_{ij} = \mathcal{A}^{\times}_{ij}    -\frac{1}{2}(\vec{\gamma}^{P}\cdot\vec{y})\mathcal{A}^{x}_{ij} -\frac{1}{2}(\vec{\gamma}^{P}\cdot\vec{x})\mathcal{A}^{y}_{ij}
\end{equation}
where $\mathcal{A}^{x}_{ij} = x_{i}k_{j}+k_{i}x_{j}$ and $\mathcal{A}^{y}_{ij} = y_{i}k_{j}+k_{i}y_{j}$ are the polarization tensors for the vector -x and vector- y polarizations respectively.

 It is evident from the above expressions \eqref{rotated_plus_P} and \eqref{rotated_cross_P} that the polarization plane is rotated in the geometric optics limit in a similar manner to what we have observed in GR \cite{Hou:2019wdg}. However, the rotation is different from GR as a consequence of the extra conformal factor $f'(\hat{R})$ appearing in $\vec{\gamma}^{P}$. Let us recall that it is customary in the literature to neglect the spin 2 nature of gravitational waves and it is acceptable as the polarization tensor is parallel transported along the geodesic as we saw in section \ref{go}.

\subsection{Lensing by Singular Isothermal Sphere }

Singular Isothermal Sphere (SIS) \cite{Kormann1994} is one of the simplest parametrizations of the axially symmetric lens model where the matter content of the lens behaves like an ideal gas in thermal and hydrostatic equilibrium.
The SIS density profile is given by the density distribution
\begin{equation}
    \rho(r) = \frac{\sigma_{v}^{2}}{2\pi G r^{2}}
\end{equation}
where $\sigma_{v}$ is the velocity distribution of the gas particles. 
Using the above density distribution and the Poisson equation given in \eqref{poisson} it is easy to 
calculate the convergence 
\begin{eqnarray}
    \kappa({\vec{\xi}}) &=& \frac{1}{c^{2}}\frac{D_{L}D_{LS}}{D_{S}}\int_{-\infty}^{\infty}\nabla^{2}\Phi(\vec{\xi},z) dz \nonumber \\
    &=&  \frac{4\pi G}{c^{2}}\frac{D_{L}D_{LS}}{D_{S}}\int_{-\infty}^{\infty} (\rho + 2\alpha \nabla^{2}\rho)dz 
\end{eqnarray}
which provides
\begin{equation}
     \kappa({\vec{\xi}}) = \frac{2\pi\sigma_{v}^{2}}{c^{2}}\frac{D_{L}D_{LS}}{D_{S}}\Big[  \frac{1}{\xi} + \frac{2\alpha}{\xi^{3}}\Big].
\end{equation}

The above equation can be re-written in terms of the dimensionless variable $x = \xi/\xi_{0}$,
\begin{equation}\label{main3}
\kappa(x) = \frac{1}{2x} + \frac{\alpha}{x^{3}\xi_{0}^{2}} \,,
\end{equation}
where the length scale $\xi_0$ and is given by 
\begin{equation}
    \xi_{0} = 4\pi \frac{\sigma_{v}^{2}}{c^{2}}\frac{D_{L}D_{LS}}{D_{S}}.
\end{equation}
As expected, taking $\alpha =0$ recovers the convergence in GR for the SIS density profile. 

Lensing potential $\Psi(\vec{x}) = \Psi(x)$ can be obtained from the 2D Poisson equation $\nabla^2_{\vec{x}} \Psi(x) = 2 \kappa(x).$ For this purpose it is convenient to introduce polar coordinates in the lens plane: $\vec{x} = x(\cos{\varphi}, \sin{\varphi})$. Then $\nabla^2_{\vec{x}} = \frac{1}{x}\frac{\partial}{\partial x}\left(x \frac{\partial}{\partial x}\right) + \frac{1}{x^2} \frac{\partial^2}{\partial \varphi^2}$ and 2D Poisson equation reads
\begin{equation}
   \frac{1}{x}\frac{d}{dx}\left(x \frac{d \Psi}{dx}\right) = \frac{1}{x} + \frac{2 \alpha}{x^{3}\xi_{0}^{2}}.
\end{equation}
Its solution is
\begin{equation}\label{main2}
    \Psi(x) = x + \frac{2 \alpha}{\xi_0^2 x} \,.
\end{equation}
Taking the derivative of the lensing potential \eqref{lensang} gives us the corresponding deflection angle $\vec \gamma$ for the Palatini polynomial $f(\hat{R})$ models,
\begin{equation}\label{def_angle_P}
   \vec \gamma^{P}(x) = \frac{\vec{x}}{x} - \frac{2 \alpha}{\xi_0^2} \frac{\vec{x}}{x^{3}} \,.
\end{equation}
In addition to the convergence $\kappa(x)$ being informative about Ricci part of curvature effect on the bundle of null geodesics, shear modulus\footnote{For more details see Appendix \ref{appLensingJacobian}.} $\gamma(x)$ quantifies the magnitude of the Weyl part of the curvature tensor. In the case of Palatini's theory one has, for the SIS lens: 
\begin{align}
    \gamma = \frac{1}{2 x^3} \left|\frac{6 \alpha}{\xi_0^2} - x^2 \right|
\end{align}
and the magnification is
\begin{align}
    \mu = x^6 \left(x^3 - \frac{4 \alpha}{\xi_0^2} \right)^{-1} \left(x^3 - x^2 + \frac{2 \alpha}{\xi_0^2} \right)^{-1} \,,
\end{align}
where again GR formulae are recovered in the limit of vanishing $\alpha$.

\section{Conclusions} \label{conclusion}

The aim of the present research was to examine the effects of Palatini $f(\hat R)$ gravity on the propagation of gravitational waves and gravitational lensing of them, in particular. To our knowledge, this is the first work in which the effects of metric-affine gravity were studied in the context of the lensing of gravitational waves. Firstly, we have confirmed the results obtained by previous works in the literature \cite{jimenez2015tensor,jimenez2018born}, demonstrating that gravitational waves propagate along autoparallel curves in this theory which coincides with geodesics in general relativity. This consistency reinforces the validity of our approach and contributes to the overall understanding of gravitational waves in modified theories of gravity. Following the standard procedure, we have studied using WKB approximation the geometric optics limit in Palatini $f(\hat{R})$ gravity. This allowed us to successfully compute the evolution of gravitational wave amplitude and explore the effects of metric-affine gravity on the rotation of the gravitational wave polarization plane. Furthermore, as a useful and realistic example,  we discussed the SIS model, which is a robust model for a galaxy acting as a lens, in the context of  Palatini $f(\hat{R})$ gravity.

The examined model of gravity reduces to GR in a vacuum, it does not provide any changes in the gravitational wave amplitude and other lensing properties for a point lens. Therefore, one has to consider the propagation of GWs through the region of space-time where a non-zero energy density distribution is present or in a close vicinity of it. Note that the non-relativistic gravitational potential describing such a situation carries an additional term due to the modification of the Poisson equation \eqref{poisson}. An example of the first case is when a GW signal emitted from the source travels not exactly through the vacuum but rather through the dark matter halo of the host galaxy, through the dark matter of the galaxy cluster to which the host belongs, and through the dark matter halo of our Galaxy.  
This attenuates the GW amplitude in the metric-affine gravity theory studied. Moreover, we have shown that the GW amplitude evolution equation can be expressed as a correction to the measured luminosity distance \eqref{luminosity_distance_corr} of the source. One should note that the corrections depend on the matter density distribution along the line of sight through the conformal factor. As an example, we considered a galaxy halo modeled by NFW (SIS) density profile. Taking the value of $\alpha \sim 10^{9}$ m$^{2}$ we found that the corrections are of the order of $10^{-42} \, (10^{-43})$, which is negligible. However, the corrections could be significant -- of order of $10^{-2}$ if one uses the constraints for $\alpha \sim 10^{49}$ m$^{2}$ obtained from SNIa and BAO observations \cite{Gomes:2023xzk}. Furthermore, we have derived the corrections to the polarization tensor \eqref{rotated_plus_P} and \eqref{rotated_cross_P}  and found that the polarization plane is rotated and a vector polarization appears. However, just like in GR this is illusory and can be removed by proper gauge choice. It is worth mentioning that these corrections depend on the modified deflection angle \eqref{def_angle_P} and hence they are theory dependent. As an example of the second case, we studied the SIS lens model and we demonstrated that Palatini $f(\hat{R})$ gravity indeed modifies the convergence \eqref{main3}, lensing potential \eqref{main2}, and the deflection angle \eqref{def_angle_P}, reducing to the GR limit when $\alpha$ vanishes. Let us stress that the exemplary value of $\alpha$ used by us is not the firmly established one and can be updated in the future. Hence the claims of negligible modified gravity effects are not firmly established, as well.   Therefore, the results we obtained could be valuable for further metric-affine gravity vs.\ GR tests involving lensing of GWs and comparison of luminosity distances measured from electromagnetic and GW sources.

\subsection*{Acknowledgements}
This work was supported by the European Regional Development Fund through the Center of Excellence TK133 ``The Dark Side of the Universe." SH acknowledges the Dora Plus grant for visiting students supported by \textit{European Regional Development Fund} and Laboratory of Theoretical Physics, University of Tartu where this research is carried out. AW acknowledges financial support from MICINN (Spain) {\it Ayuda Juan de la Cierva - incorporaci\'on} 2020 No. IJC2020-044751-I. LJ and MS were supported by the Estonian Research Council grant PRG356 ``Gauge Gravity'' and  TK202 ``Foundations of the Universe.''

\bibliographystyle{utphys}    % VERY NICE STYLE
\bibliography{GW_literature}

\providecommand{\href}[2]{#2}\begingroup\raggedright\begin{thebibliography}{100}

\bibitem{LIGOScientific:2016aoc}
{\bf LIGO Scientific, Virgo} Collaboration, B.~P. Abbott {\em et al.},
  ``{Observation of Gravitational Waves from a Binary Black Hole Merger},''
  \href{http://dx.doi.org/10.1103/PhysRevLett.116.061102}{{\em Phys. Rev.
  Lett.} {\bf 116} (2016) no.~6, 061102},
  \href{http://arxiv.org/abs/1602.03837}{{\tt arXiv:1602.03837 [gr-qc]}}.

\bibitem{LIGOScientific:2021djp}
{\bf LIGO Scientific, VIRGO, KAGRA} Collaboration, R.~Abbott {\em et al.},
  ``{GWTC-3: Compact Binary Coalescences Observed by LIGO and Virgo During the
  Second Part of the Third Observing Run},''
  \href{http://arxiv.org/abs/2111.03606}{{\tt arXiv:2111.03606 [gr-qc]}}.

\bibitem{Antoniadis:2023rey}
J.~Antoniadis {\em et al.}, ``{The second data release from the European Pulsar
  Timing Array III. Search for gravitational wave signals},''
  \href{http://arxiv.org/abs/2306.16214}{{\tt arXiv:2306.16214 [astro-ph.HE]}}.

\bibitem{NANOGrav:2023gor}
{\bf NANOGrav} Collaboration, G.~Agazie {\em et al.}, ``{The NANOGrav 15 yr
  Data Set: Evidence for a Gravitational-wave Background},''
  \href{http://dx.doi.org/10.3847/2041-8213/acdac6}{{\em Astrophys. J. Lett.}
  {\bf 951} (2023) no.~1, L8}, \href{http://arxiv.org/abs/2306.16213}{{\tt
  arXiv:2306.16213 [astro-ph.HE]}}.

\bibitem{Xu:2023wog}
H.~Xu {\em et al.}, ``{Searching for the Nano-Hertz Stochastic Gravitational
  Wave Background with the Chinese Pulsar Timing Array Data Release I},''
  \href{http://dx.doi.org/10.1088/1674-4527/acdfa5}{{\em Res. Astron.
  Astrophys.} {\bf 23} (2023) no.~7, 075024},
  \href{http://arxiv.org/abs/2306.16216}{{\tt arXiv:2306.16216 [astro-ph.HE]}}.

\bibitem{Reardon:2023gzh}
D.~J. Reardon {\em et al.}, ``{Search for an Isotropic Gravitational-wave
  Background with the Parkes Pulsar Timing Array},''
  \href{http://dx.doi.org/10.3847/2041-8213/acdd02}{{\em Astrophys. J. Lett.}
  {\bf 951} (2023) no.~1, L6}, \href{http://arxiv.org/abs/2306.16215}{{\tt
  arXiv:2306.16215 [astro-ph.HE]}}.

\bibitem{scalartensor}
J.~Sakstein and B.~Jain, ``{Implications of the Neutron Star Merger GW170817
  for Cosmological Scalar-Tensor Theories},''
  \href{http://dx.doi.org/10.1103/PhysRevLett.119.251303}{{\em Phys. Rev.
  Lett.} {\bf 119} (2017) no.~25, 251303},
  \href{http://arxiv.org/abs/1710.05893}{{\tt arXiv:1710.05893 [astro-ph.CO]}}.

\bibitem{dead_ends}
J.~M. Ezquiaga and M.~Zumalac\'arregui, ``{Dark Energy After GW170817: Dead
  Ends and the Road Ahead},''
  \href{http://dx.doi.org/10.1103/PhysRevLett.119.251304}{{\em Phys. Rev.
  Lett.} {\bf 119} (2017) no.~25, 251304},
  \href{http://arxiv.org/abs/1710.05901}{{\tt arXiv:1710.05901 [astro-ph.CO]}}.

\bibitem{Tessa_test}
T.~Baker, E.~Bellini, P.~G. Ferreira, M.~Lagos, J.~Noller, and I.~Sawicki,
  ``{Strong constraints on cosmological gravity from GW170817 and GRB
  170817A},'' \href{http://dx.doi.org/10.1103/PhysRevLett.119.251301}{{\em
  Phys. Rev. Lett.} {\bf 119} (2017) no.~25, 251301},
  \href{http://arxiv.org/abs/1710.06394}{{\tt arXiv:1710.06394 [astro-ph.CO]}}.

\bibitem{Gravitational_lenses1992}
P.~{Schneider}, J.~{Ehlers}, and E.~E. {Falco},
  \href{http://dx.doi.org/10.1007/978-3-662-03758-4}{{\em {Gravitational
  Lenses}}}.
\newblock Astronomy and Astrophysics Library. Springer, 1992.

\bibitem{Hannuksela:2019kle}
O.~A. Hannuksela, K.~Haris, K.~K.~Y. Ng, S.~Kumar, A.~K. Mehta, D.~Keitel,
  T.~G.~F. Li, and P.~Ajith, ``{Search for gravitational lensing signatures in
  LIGO-Virgo binary black hole events},''
  \href{http://dx.doi.org/10.3847/2041-8213/ab0c0f}{{\em Astrophys. J. Lett.}
  {\bf 874} (2019) no.~1, L2}, \href{http://arxiv.org/abs/1901.02674}{{\tt
  arXiv:1901.02674 [gr-qc]}}.

\bibitem{LIGOScientific:2021izm}
{\bf LIGO Scientific, VIRGO} Collaboration, R.~Abbott {\em et al.}, ``{Search
  for Lensing Signatures in the Gravitational-Wave Observations from the First
  Half of LIGO\textendash{}Virgo\textquoteright{}s Third Observing Run},''
  \href{http://dx.doi.org/10.3847/1538-4357/ac23db}{{\em Astrophys. J.} {\bf
  923} (2021) no.~1, 14}, \href{http://arxiv.org/abs/2105.06384}{{\tt
  arXiv:2105.06384 [gr-qc]}}.

\bibitem{LIGOScientific:2023bwz}
{\bf LIGO Scientific, VIRGO, KAGRA} Collaboration, R.~Abbott {\em et al.},
  ``{Search for gravitational-lensing signatures in the full third observing
  run of the LIGO-Virgo network},'' \href{http://arxiv.org/abs/2304.08393}{{\tt
  arXiv:2304.08393 [gr-qc]}}.

\bibitem{Sereno:2011ty}
M.~Sereno, P.~Jetzer, A.~Sesana, and M.~Volonteri, ``{Cosmography with strong
  lensing of LISA gravitational wave sources},''
  \href{http://dx.doi.org/10.1111/j.1365-2966.2011.18895.x}{{\em Mon. Not. Roy.
  Astron. Soc.} {\bf 415} (2011)  2773},
  \href{http://arxiv.org/abs/1104.1977}{{\tt arXiv:1104.1977 [astro-ph.CO]}}.

\bibitem{Cao:2019kgn}
S.~Cao, J.~Qi, Z.~Cao, M.~Biesiada, J.~Li, Y.~Pan, and Z.-H. Zhu, ``{Direct
  test of the FLRW metric from strongly lensed gravitational wave
  observations},'' \href{http://dx.doi.org/10.1038/s41598-019-47616-4}{{\em
  Sci. Rep.} {\bf 9} (2019) no.~1, 11608},
  \href{http://arxiv.org/abs/1910.10365}{{\tt arXiv:1910.10365 [astro-ph.CO]}}.

\bibitem{Liao:2017ioi}
K.~Liao, X.-L. Fan, X.-H. Ding, M.~Biesiada, and Z.-H. Zhu, ``{Precision
  cosmology from future lensed gravitational wave and electromagnetic
  signals},'' \href{http://dx.doi.org/10.1038/s41467-017-01152-9}{{\em Nature
  Commun.} {\bf 8} (2017) no.~1, 1148},
  \href{http://arxiv.org/abs/1703.04151}{{\tt arXiv:1703.04151 [astro-ph.CO]}}.
  [Erratum: Nature Commun. 8, 2136 (2017)].

\bibitem{Li:2019rns}
Y.~Li, X.~Fan, and L.~Gou, ``{Constraining Cosmological Parameters in the FLRW
  Metric with Lensed GW+EM Signals},''
  \href{http://dx.doi.org/10.3847/1538-4357/ab037e}{{\em Astrophys. J.} {\bf
  873} (2019) no.~1, 37}, \href{http://arxiv.org/abs/1901.10638}{{\tt
  arXiv:1901.10638 [astro-ph.CO]}}.

\bibitem{Biesiada:2021pzo}
M.~Biesiada and S.~Harikumar, ``{Gravitational Lensing of Continuous
  Gravitational Waves},'' \href{http://dx.doi.org/10.3390/universe7120502}{{\em
  Universe} {\bf 7} (2021) no.~12, 502},
  \href{http://arxiv.org/abs/2111.05963}{{\tt arXiv:2111.05963 [gr-qc]}}.

\bibitem{Grespan:2023cpa}
M.~Grespan and M.~Biesiada, ``{Strong Gravitational Lensing of Gravitational
  Waves: A Review},'' \href{http://dx.doi.org/10.3390/universe9050200}{{\em
  Universe} {\bf 9} (2023) no.~5, 200}.

\bibitem{Lai:2018rto}
K.-H. Lai, O.~A. Hannuksela, A.~Herrera-Mart\'\i{}n, J.~M. Diego,
  T.~Broadhurst, and T.~G.~F. Li, ``{Discovering intermediate-mass black hole
  lenses through gravitational wave lensing},''
  \href{http://dx.doi.org/10.1103/PhysRevD.98.083005}{{\em Phys. Rev. D} {\bf
  98} (2018) no.~8, 083005}, \href{http://arxiv.org/abs/1801.07840}{{\tt
  arXiv:1801.07840 [gr-qc]}}.

\bibitem{meena2023gravitational}
A.~K. Meena, ``Gravitational lensing of gravitational waves: Probing
  intermediate mass black holes in galaxy lenses with global minima,''
  \href{http://arxiv.org/abs/2305.02880}{{\tt arXiv:2305.02880 [astro-ph.CO]}}.

\bibitem{Diego:2019rzc}
J.~M. Diego, ``{Constraining the abundance of primordial black holes with
  gravitational lensing of gravitational waves at LIGO frequencies},''
  \href{http://dx.doi.org/10.1103/PhysRevD.101.123512}{{\em Phys. Rev. D} {\bf
  101} (2020) no.~12, 123512}, \href{http://arxiv.org/abs/1911.05736}{{\tt
  arXiv:1911.05736 [astro-ph.CO]}}.

\bibitem{Oguri:2020ldf}
M.~Oguri and R.~Takahashi, ``{Probing Dark Low-mass Halos and Primordial Black
  Holes with Frequency-dependent Gravitational Lensing Dispersions of
  Gravitational Waves},''
  \href{http://dx.doi.org/10.3847/1538-4357/abafab}{{\em Astrophys. J.} {\bf
  901} (2020) no.~1, 58}, \href{http://arxiv.org/abs/2007.01936}{{\tt
  arXiv:2007.01936 [astro-ph.CO]}}.

\bibitem{Ezquiaga:2020dao}
J.~M. Ezquiaga and M.~Zumalac\'arregui, ``{Gravitational wave lensing beyond
  general relativity: birefringence, echoes and shadows},''
  \href{http://dx.doi.org/10.1103/PhysRevD.102.124048}{{\em Phys. Rev. D} {\bf
  102} (2020) no.~12, 124048}, \href{http://arxiv.org/abs/2009.12187}{{\tt
  arXiv:2009.12187 [gr-qc]}}.

\bibitem{Goyal:2020bkm}
S.~Goyal, K.~Haris, A.~K. Mehta, and P.~Ajith, ``{Testing the nature of
  gravitational-wave polarizations using strongly lensed signals},''
  \href{http://dx.doi.org/10.1103/PhysRevD.103.024038}{{\em Phys. Rev. D} {\bf
  103} (2021) no.~2, 024038}, \href{http://arxiv.org/abs/2008.07060}{{\tt
  arXiv:2008.07060 [gr-qc]}}.

\bibitem{Fan:2016swi}
X.-L. Fan, K.~Liao, M.~Biesiada, A.~Piorkowska-Kurpas, and Z.-H. Zhu, ``{Speed
  of Gravitational Waves from Strongly Lensed Gravitational Waves and
  Electromagnetic Signals},''
  \href{http://dx.doi.org/10.1103/PhysRevLett.118.091102}{{\em Phys. Rev.
  Lett.} {\bf 118} (2017) no.~9, 091102},
  \href{http://arxiv.org/abs/1612.04095}{{\tt arXiv:1612.04095 [gr-qc]}}.

\bibitem{Collett:2016dey}
T.~E. Collett and D.~Bacon, ``{Testing the speed of gravitational waves over
  cosmological distances with strong gravitational lensing},''
  \href{http://dx.doi.org/10.1103/PhysRevLett.118.091101}{{\em Phys. Rev.
  Lett.} {\bf 118} (2017) no.~9, 091101},
  \href{http://arxiv.org/abs/1602.05882}{{\tt arXiv:1602.05882 [astro-ph.HE]}}.

\bibitem{Sharma:2023vme}
V.~K. Sharma, S.~Harikumar, M.~Grespan, M.~Biesiada, and M.~M. Verma,
  ``{Probing massive gravitons in $f(R)$ with lensed gravitational waves},''
  \href{http://arxiv.org/abs/2310.10346}{{\tt arXiv:2310.10346 [gr-qc]}}.

\bibitem{Mishra:2023vzo}
A.~Mishra, N.~V. Krishnendu, and A.~Ganguly, ``{Unveiling Microlensing Biases
  in Testing General Relativity with Gravitational Waves},''
  \href{http://arxiv.org/abs/2311.08446}{{\tt arXiv:2311.08446 [gr-qc]}}.

\bibitem{Baker:2016reh}
T.~Baker and M.~Trodden, ``{Multimessenger time delays from lensed
  gravitational waves},''
  \href{http://dx.doi.org/10.1103/PhysRevD.95.063512}{{\em Phys. Rev. D} {\bf
  95} (2017) no.~6, 063512}, \href{http://arxiv.org/abs/1612.02004}{{\tt
  arXiv:1612.02004 [astro-ph.CO]}}.

\bibitem{Maggiore:2019uih}
M.~Maggiore {\em et al.}, ``{Science Case for the Einstein Telescope},''
  \href{http://dx.doi.org/10.1088/1475-7516/2020/03/050}{{\em JCAP} {\bf 03}
  (2020)  050}, \href{http://arxiv.org/abs/1912.02622}{{\tt arXiv:1912.02622
  [astro-ph.CO]}}.

\bibitem{Evans:2021gyd}
M.~Evans {\em et al.}, ``{A Horizon Study for Cosmic Explorer: Science,
  Observatories, and Community},'' \href{http://arxiv.org/abs/2109.09882}{{\tt
  arXiv:2109.09882 [astro-ph.IM]}}.

\bibitem{Kawamura:2020pcg}
S.~Kawamura {\em et al.}, ``{Current status of space gravitational wave antenna
  DECIGO and B-DECIGO},'' \href{http://dx.doi.org/10.1093/ptep/ptab019}{{\em
  PTEP} {\bf 2021} (2021) no.~5, 05A105},
  \href{http://arxiv.org/abs/2006.13545}{{\tt arXiv:2006.13545 [gr-qc]}}.

\bibitem{lisa_2017}
P.~A.-S. et~al., ``Laser interferometer space antenna,''
  \href{http://arxiv.org/abs/1702.00786}{{\tt arXiv:1702.00786 [astro-ph.IM]}}.

\bibitem{Luo:2021qji}
Z.~Luo, Y.~Wang, Y.~Wu, W.~Hu, and G.~Jin, ``{The Taiji program: A concise
  overview},'' \href{http://dx.doi.org/10.1093/ptep/ptaa083}{{\em PTEP} {\bf
  2021} (2021) no.~5, 05A108}.

\bibitem{TianQin:2015yph}
{\bf TianQin} Collaboration, J.~Luo {\em et al.}, ``{TianQin: a space-borne
  gravitational wave detector},''
  \href{http://dx.doi.org/10.1088/0264-9381/33/3/035010}{{\em Class. Quant.
  Grav.} {\bf 33} (2016) no.~3, 035010},
  \href{http://arxiv.org/abs/1512.02076}{{\tt arXiv:1512.02076 [astro-ph.IM]}}.

\bibitem{Li:2018prc}
S.-S. Li, S.~Mao, Y.~Zhao, and Y.~Lu, ``{Gravitational lensing of gravitational
  waves: A statistical perspective},''
  \href{http://dx.doi.org/10.1093/mnras/sty411}{{\em Mon. Not. Roy. Astron.
  Soc.} {\bf 476} (2018) no.~2, 2220--2229},
  \href{http://arxiv.org/abs/1802.05089}{{\tt arXiv:1802.05089 [astro-ph.CO]}}.

\bibitem{Biesiada:2014kwa}
M.~Biesiada, X.~Ding, A.~Piorkowska, and Z.-H. Zhu, ``{Strong gravitational
  lensing of gravitational waves from double compact binaries - perspectives
  for the Einstein Telescope},''
  \href{http://dx.doi.org/10.1088/1475-7516/2014/10/080}{{\em JCAP} {\bf 10}
  (2014)  080}, \href{http://arxiv.org/abs/1409.8360}{{\tt arXiv:1409.8360
  [astro-ph.HE]}}.

\bibitem{Yang_2019}
L.~Yang, X.~Ding, M.~Biesiada, K.~Liao, and Z.-H. Zhu, ``{How Does the
  Earth\textquoteright{}s Rotation Affect Predictions of Gravitational Wave
  Strong Lensing Rates?},''
  \href{http://dx.doi.org/10.3847/1538-4357/ab095c}{{\em Astrophys. J.} {\bf
  874} (2019) no.~2, 139}, \href{http://arxiv.org/abs/1903.11079}{{\tt
  arXiv:1903.11079 [astro-ph.GA]}}.

\bibitem{Lilan2021}
L.~Yang, S.~Wu, K.~Liao, X.~Ding, Z.~You, Z.~Cao, M.~Biesiada, and Z.-H. Zhu,
  ``{Event rate predictions of strongly lensed gravitational waves with
  detector networks and more realistic templates},''
  \href{http://dx.doi.org/10.1093/mnras/stab3298}{{\em Mon. Not. Roy. Astron.
  Soc.} {\bf 509} (2021) no.~3, 3772--3778},
  \href{http://arxiv.org/abs/2105.07011}{{\tt arXiv:2105.07011 [astro-ph.GA]}}.

\bibitem{Piorkowska2021}
A.~Pi\'orkowska-Kurpas, S.~Hou, M.~Biesiada, X.~Ding, S.~Cao, X.~Fan,
  S.~Kawamura, and Z.-H. Zhu, ``{Inspiraling Double Compact Object Detection
  and Lensing Rate: Forecast for DECIGO and B-DECIGO},''
  \href{http://dx.doi.org/10.3847/1538-4357/abd482}{{\em Astrophys. J.} {\bf
  908} (2021) no.~2, 196}, \href{http://arxiv.org/abs/2005.08727}{{\tt
  arXiv:2005.08727 [astro-ph.HE]}}.

\bibitem{Nakamura:1999uwi}
T.~T. Nakamura and S.~Deguchi, ``{Wave Optics in Gravitational Lensing},''
  \href{http://dx.doi.org/10.1143/ptps.133.137}{{\em Prog. Theor. Phys. Suppl.}
  {\bf 133} (1999)  137--153}.

\bibitem{Takahashi:2003ix}
R.~Takahashi and T.~Nakamura, ``{Wave effects in gravitational lensing of
  gravitational waves from chirping binaries},''
  \href{http://dx.doi.org/10.1086/377430}{{\em Astrophys. J.} {\bf 595} (2003)
  1039--1051}, \href{http://arxiv.org/abs/astro-ph/0305055}{{\tt
  arXiv:astro-ph/0305055}}.

\bibitem{Will_book}
C.~M. Will, \href{http://dx.doi.org/10.1017/CBO9780511564246}{{\em {Theory and
  Experiment in Gravitational Physics}}}.
\newblock Cambridge University Press, 1993.

\bibitem{Eardley73}
D.~M. Eardley, D.~L. Lee, A.~P. Lightman, R.~V. Wagoner, and C.~M. Will,
  ``{Gravitational-wave observations as a tool for testing relativistic
  gravity},'' \href{http://dx.doi.org/10.1103/PhysRevLett.30.884}{{\em Phys.
  Rev. Lett.} {\bf 30} (1973)  884--886}.

\bibitem{Thorne73}
K.~S. Thorne, D.~L. Lee, and A.~P. Lightman, ``{Foundations for a Theory of
  Gravitation Theories},''
  \href{http://dx.doi.org/10.1103/PhysRevD.7.3563}{{\em Phys. Rev. D} {\bf 7}
  (1973)  3563--3578}.

\bibitem{Hou:2019wdg}
S.~Hou, X.-L. Fan, and Z.-H. Zhu, ``{Gravitational Lensing of Gravitational
  Waves: Rotation of Polarization Plane},''
  \href{http://dx.doi.org/10.1103/PhysRevD.100.064028}{{\em Phys. Rev. D} {\bf
  100} (2019) no.~6, 064028}, \href{http://arxiv.org/abs/1907.07486}{{\tt
  arXiv:1907.07486 [gr-qc]}}.

\bibitem{Dalang:2021qhu}
C.~Dalang, G.~Cusin, and M.~Lagos, ``{Polarization distortions of lensed
  gravitational waves},''
  \href{http://dx.doi.org/10.1103/PhysRevD.105.024005}{{\em Phys. Rev. D} {\bf
  105} (2022) no.~2, 024005}, \href{http://arxiv.org/abs/2104.10119}{{\tt
  arXiv:2104.10119 [gr-qc]}}.

\bibitem{Cusin:2019rmt}
G.~Cusin and M.~Lagos, ``{Gravitational wave propagation beyond geometric
  optics},'' \href{http://dx.doi.org/10.1103/PhysRevD.101.044041}{{\em Phys.
  Rev. D} {\bf 101} (2020) no.~4, 044041},
  \href{http://arxiv.org/abs/1910.13326}{{\tt arXiv:1910.13326 [gr-qc]}}.

\bibitem{Olmo:2011uz}
G.~J. Olmo, ``{Palatini Approach to Modified Gravity: f(R) Theories and
  Beyond},'' \href{http://dx.doi.org/10.1142/S0218271811018925}{{\em Int. J.
  Mod. Phys. D} {\bf 20} (2011)  413--462},
  \href{http://arxiv.org/abs/1101.3864}{{\tt arXiv:1101.3864 [gr-qc]}}.

\bibitem{Sotiriou:2006sf}
T.~P. Sotiriou, ``{Curvature scalar instability in f(R) gravity},''
  \href{http://dx.doi.org/10.1016/j.physletb.2007.01.003}{{\em Phys. Lett. B}
  {\bf 645} (2007)  389--392}, \href{http://arxiv.org/abs/gr-qc/0611107}{{\tt
  arXiv:gr-qc/0611107}}.

\bibitem{Stachowski:2016zio}
A.~Stachowski, M.~Szyd\l{}owski, and A.~Borowiec, ``{Starobinsky cosmological
  model in Palatini formalism},''
  \href{http://dx.doi.org/10.1140/epjc/s10052-017-4981-8}{{\em Eur. Phys. J. C}
  {\bf 77} (2017) no.~6, 406}, \href{http://arxiv.org/abs/1608.03196}{{\tt
  arXiv:1608.03196 [gr-qc]}}.

\bibitem{Szydlowski:2017uuy}
M.~Szyd\l{}owski, A.~Stachowski, and A.~Borowiec, ``{Emergence of running dark
  energy from polynomial f(R) theory in Palatini formalism},''
  \href{http://dx.doi.org/10.1140/epjc/s10052-017-5181-2}{{\em Eur. Phys. J. C}
  {\bf 77} (2017) no.~9, 603}, \href{http://arxiv.org/abs/1707.01948}{{\tt
  arXiv:1707.01948 [gr-qc]}}.

\bibitem{TeppaPannia:2018ale}
F.~A. Teppa~Pannia, S.~E. Perez~Bergliaffa, and N.~Manske, ``{Cosmography and
  the redshift drift in Palatini $f({\cal R})$ theories},''
  \href{http://dx.doi.org/10.1140/epjc/s10052-019-6764-x}{{\em Eur. Phys. J. C}
  {\bf 79} (2019) no.~3, 267}, \href{http://arxiv.org/abs/1811.08176}{{\tt
  arXiv:1811.08176 [gr-qc]}}.

\bibitem{Pinto:2018rfg}
P.~Pinto, L.~Del~Vecchio, L.~Fatibene, and M.~Ferraris, ``{Extended cosmology
  in Palatini f(R)-theories},''
  \href{http://dx.doi.org/10.1088/1475-7516/2018/11/044}{{\em JCAP} {\bf 11}
  (2018)  044}, \href{http://arxiv.org/abs/1807.00397}{{\tt arXiv:1807.00397
  [gr-qc]}}.

\bibitem{Camera:2022myt}
S.~Camera, S.~Capozziello, L.~Fatibene, and A.~Orizzonte, ``{The effective
  equation of state in Palatini $f({{\mathcal {R}}})$ cosmology},''
  \href{http://dx.doi.org/10.1140/epjp/s13360-023-03676-0}{{\em Eur. Phys. J.
  Plus} {\bf 138} (2023) no.~2, 180},
  \href{http://arxiv.org/abs/2212.13825}{{\tt arXiv:2212.13825 [gr-qc]}}.

\bibitem{Gialamas:2023flv}
I.~D. Gialamas, A.~Karam, T.~D. Pappas, and E.~Tomberg, ``{Implications of
  Palatini gravity for inflation and beyond},''
  \href{http://arxiv.org/abs/2303.14148}{{\tt arXiv:2303.14148 [gr-qc]}}.

\bibitem{Olmo:2019qsj}
G.~J. Olmo, D.~Rubiera-Garcia, and A.~Wojnar, ``{Minimum main sequence mass in
  quadratic Palatini $f(R)$ gravity},''
  \href{http://dx.doi.org/10.1103/PhysRevD.100.044020}{{\em Phys. Rev. D} {\bf
  100} (2019) no.~4, 044020}, \href{http://arxiv.org/abs/1906.04629}{{\tt
  arXiv:1906.04629 [gr-qc]}}.

\bibitem{Olmo:2019flu}
G.~J. Olmo, D.~Rubiera-Garcia, and A.~Wojnar, ``{Stellar structure models in
  modified theories of gravity: Lessons and challenges},''
  \href{http://dx.doi.org/10.1016/j.physrep.2020.07.001}{{\em Phys. Rept.} {\bf
  876} (2020)  1--75}, \href{http://arxiv.org/abs/1912.05202}{{\tt
  arXiv:1912.05202 [gr-qc]}}.

\bibitem{Wojnar:2020txr}
A.~Wojnar, ``{Early evolutionary tracks of low-mass stellar objects in modified
  gravity},'' \href{http://dx.doi.org/10.1103/PhysRevD.102.124045}{{\em Phys.
  Rev. D} {\bf 102} (2020) no.~12, 124045},
  \href{http://arxiv.org/abs/2007.13451}{{\tt arXiv:2007.13451 [gr-qc]}}.

\bibitem{Wojnar:2020frr}
A.~Wojnar, ``{Lithium abundance is a gravitational model dependent quantity},''
  \href{http://dx.doi.org/10.1103/PhysRevD.103.044037}{{\em Phys. Rev. D} {\bf
  103} (2021) no.~4, 044037}, \href{http://arxiv.org/abs/2009.10983}{{\tt
  arXiv:2009.10983 [gr-qc]}}.

\bibitem{Benito:2021ywe}
M.~Benito and A.~Wojnar, ``{Cooling process of brown dwarfs in Palatini f(R)
  gravity},'' \href{http://dx.doi.org/10.1103/PhysRevD.103.064032}{{\em Phys.
  Rev. D} {\bf 103} (2021) no.~6, 064032},
  \href{http://arxiv.org/abs/2101.02146}{{\tt arXiv:2101.02146 [gr-qc]}}.

\bibitem{Wojnar:2021xbr}
A.~Wojnar, ``{Jupiter and jovian exoplanets in Palatini f(R\textasciimacron{})
  gravity},'' \href{http://dx.doi.org/10.1103/PhysRevD.104.104058}{{\em Phys.
  Rev. D} {\bf 104} (2021) no.~10, 104058},
  \href{http://arxiv.org/abs/2108.13528}{{\tt arXiv:2108.13528 [gr-qc]}}.

\bibitem{Kozak:2021ghd}
A.~Kozak and A.~Wojnar, ``{Metric-affine gravity effects on terrestrial
  exoplanet profiles},''
  \href{http://dx.doi.org/10.1103/PhysRevD.104.084097}{{\em Phys. Rev. D} {\bf
  104} (2021) no.~8, 084097}, \href{http://arxiv.org/abs/2106.14219}{{\tt
  arXiv:2106.14219 [gr-qc]}}.

\bibitem{Sarmah:2021ule}
L.~Sarmah, S.~Kalita, and A.~Wojnar, ``{Stability criterion for white dwarfs in
  Palatini f(R) gravity},''
  \href{http://dx.doi.org/10.1103/PhysRevD.105.024028}{{\em Phys. Rev. D} {\bf
  105} (2022) no.~2, 024028}, \href{http://arxiv.org/abs/2111.08029}{{\tt
  arXiv:2111.08029 [gr-qc]}}.

\bibitem{Kalita:2022trq}
S.~Kalita, L.~Sarmah, and A.~Wojnar, ``{Metric-affine effects in
  crystallization processes of white dwarfs},''
  \href{http://dx.doi.org/10.1103/PhysRevD.107.044072}{{\em Phys. Rev. D} {\bf
  107} (2023) no.~4, 044072}, \href{http://arxiv.org/abs/2212.04918}{{\tt
  arXiv:2212.04918 [gr-qc]}}.

\bibitem{Kozak:2023axy}
A.~Kozak and A.~Wojnar, ``{Planetary seismology as a test of modified gravity
  proposals},'' \href{http://dx.doi.org/10.1103/PhysRevD.108.044055}{{\em Phys.
  Rev. D} {\bf 108} (2023) no.~4, 044055},
  \href{http://arxiv.org/abs/2303.17213}{{\tt arXiv:2303.17213 [gr-qc]}}.

\bibitem{Toniato:2019rrd}
J.~D. Toniato, D.~C. Rodrigues, and A.~Wojnar, ``{Palatini $f(R)$ gravity in
  the solar system: post-Newtonian equations of motion and complete PPN
  parameters},'' \href{http://dx.doi.org/10.1103/PhysRevD.101.064050}{{\em
  Phys. Rev. D} {\bf 101} (2020) no.~6, 064050},
  \href{http://arxiv.org/abs/1912.12234}{{\tt arXiv:1912.12234 [gr-qc]}}.

\bibitem{Bonino:2020wps}
A.~Bonino, S.~Camera, L.~Fatibene, and A.~Orizzonte, ``{Solar System tests in
  Brans\textendash{}Dicke and Palatini $f({\mathcal {R}})$-theories},''
  \href{http://dx.doi.org/10.1140/epjp/s13360-020-00982-9}{{\em Eur. Phys. J.
  Plus} {\bf 135} (2020) no.~12, 951},
  \href{http://arxiv.org/abs/2011.06303}{{\tt arXiv:2011.06303 [gr-qc]}}.

\bibitem{Hernandez-Arboleda:2023abv}
A.~Hernandez-Arboleda, D.~C. Rodrigues, J.~D. Toniato, and A.~Wojnar,
  ``{Palatini f(R) gravity tests in weak field limit: Solar system, seismology
  and galaxies},'' \href{http://dx.doi.org/10.1142/S0219887824500282}{{\em Int.
  J. Geom. Meth. Mod. Phys.} {\bf 20} (2023) no.~Supp01, 2450028},
  \href{http://arxiv.org/abs/2306.04475}{{\tt arXiv:2306.04475 [gr-qc]}}.

\bibitem{Alves:2009eg}
M.~E.~S. Alves, O.~D. Miranda, and J.~C.~N. de~Araujo, ``{Probing the f(R)
  formalism through gravitational wave polarizations},''
  \href{http://dx.doi.org/10.1016/j.physletb.2009.08.005}{{\em Phys. Lett. B}
  {\bf 679} (2009)  401--406}, \href{http://arxiv.org/abs/0908.0861}{{\tt
  arXiv:0908.0861 [gr-qc]}}.

\bibitem{Iosifidis:2021nra}
D.~Iosifidis, ``{The Perfect Hyperfluid of Metric-Affine Gravity: The
  Foundation},'' \href{http://dx.doi.org/10.1088/1475-7516/2021/04/072}{{\em
  JCAP} {\bf 04} (2021)  072}, \href{http://arxiv.org/abs/2101.07289}{{\tt
  arXiv:2101.07289 [gr-qc]}}.

\bibitem{Koivisto:2005yk}
T.~Koivisto, ``{Covariant conservation of energy momentum in modified
  gravities},'' \href{http://dx.doi.org/10.1088/0264-9381/23/12/N01}{{\em
  Class. Quant. Grav.} {\bf 23} (2006)  4289--4296},
  \href{http://arxiv.org/abs/gr-qc/0505128}{{\tt arXiv:gr-qc/0505128}}.

\bibitem{ferraris1993universal}
M.~Ferraris, M.~Francaviglia, and I.~Volovich, ``{Universal gravitational
  equations},'' \href{http://dx.doi.org/10.1007/BF02741283}{{\em Nuovo Cim. B}
  {\bf 108} (1993)  1313--1317}.

\bibitem{ferraris1994universality}
M.~Ferraris, M.~Francaviglia, and I.~Volovich, ``{The Universality of vacuum
  Einstein equations with cosmological constant},''
  \href{http://dx.doi.org/10.1088/0264-9381/11/6/015}{{\em Class. Quant. Grav.}
  {\bf 11} (1994)  1505--1517}, \href{http://arxiv.org/abs/gr-qc/9303007}{{\tt
  arXiv:gr-qc/9303007}}.

\bibitem{schwartz2019post}
P.~K. Schwartz and D.~Giulini, ``{Post-Newtonian Hamiltonian description of an
  atom in a weak gravitational field},''
  \href{http://dx.doi.org/10.1103/PhysRevA.100.052116}{{\em Phys. Rev. A} {\bf
  100} (2019) no.~5, 052116}, \href{http://arxiv.org/abs/1908.06929}{{\tt
  arXiv:1908.06929 [quant-ph]}}.

\bibitem{olmo2008hydrogen}
G.~J. Olmo, ``{Hydrogen atom in Palatini theories of gravity},''
  \href{http://dx.doi.org/10.1103/PhysRevD.77.084021}{{\em Phys. Rev. D} {\bf
  77} (2008)  084021}, \href{http://arxiv.org/abs/0802.4038}{{\tt
  arXiv:0802.4038 [gr-qc]}}.

\bibitem{olmo2007violation}
G.~J. Olmo, ``{Violation of the Equivalence Principle in Modified Theories of
  Gravity},'' \href{http://dx.doi.org/10.1103/PhysRevLett.98.061101}{{\em Phys.
  Rev. Lett.} {\bf 98} (2007)  061101},
  \href{http://arxiv.org/abs/gr-qc/0612002}{{\tt arXiv:gr-qc/0612002}}.

\bibitem{Wojnar:2022dvo}
A.~Wojnar, ``{Fermi gas and modified gravity},''
  \href{http://dx.doi.org/10.1103/PhysRevD.107.044025}{{\em Phys. Rev. D} {\bf
  107} (2023) no.~4, 044025}, \href{http://arxiv.org/abs/2208.04023}{{\tt
  arXiv:2208.04023 [gr-qc]}}.

\bibitem{naf2009gravitational}
J.~Naf, P.~Jetzer, and M.~Sereno, ``{On Gravitational Waves in Spacetimes with
  a Nonvanishing Cosmological Constant},''
  \href{http://dx.doi.org/10.1103/PhysRevD.79.024014}{{\em Phys. Rev. D} {\bf
  79} (2009)  024014}, \href{http://arxiv.org/abs/0810.5426}{{\tt
  arXiv:0810.5426 [astro-ph]}}.

\bibitem{Olmo:2005zr}
G.~J. Olmo, ``{The Gravity Lagrangian according to solar system experiments},''
  \href{http://dx.doi.org/10.1103/PhysRevLett.95.261102}{{\em Phys. Rev. Lett.}
  {\bf 95} (2005)  261102}, \href{http://arxiv.org/abs/gr-qc/0505101}{{\tt
  arXiv:gr-qc/0505101}}.

\bibitem{Kozak:2023ruu}
A.~Kozak and A.~Wojnar, ``{Earthquakes as probing tools for gravity
  theories},'' \href{http://arxiv.org/abs/2308.01784}{{\tt arXiv:2308.01784
  [gr-qc]}}.

\bibitem{Lope-Oter:2023urz}
E.~Lope-Oter and A.~Wojnar, ``{Constraining Palatini gravity with
  GR-independent equations of state for neutron stars},''
  \href{http://dx.doi.org/10.1088/1475-7516/2024/02/017}{{\em JCAP} {\bf 02}
  (2024)  017}, \href{http://arxiv.org/abs/2306.00870}{{\tt arXiv:2306.00870
  [gr-qc]}}.

\bibitem{Wojnar:2023bvv}
A.~Wojnar, ``{Unveiling phase space modifications: A clash of modified gravity
  and the generalized uncertainty principle},''
  \href{http://dx.doi.org/10.1103/PhysRevD.109.024011}{{\em Phys. Rev. D} {\bf
  109} (2024) no.~2, 024011}, \href{http://arxiv.org/abs/2311.14066}{{\tt
  arXiv:2311.14066 [gr-qc]}}.

\bibitem{Hernandez-Arboleda:2022rim}
D.~C. Rodrigues, A.~Hernandez-Arboleda, and A.~Wojnar, ``{Normalized additional
  velocity distribution: Testing the radial profile of dark matter halos and
  MOND},'' \href{http://dx.doi.org/10.1016/j.dark.2023.101230}{{\em Phys. Dark
  Univ.} {\bf 41} (2023)  101230}, \href{http://arxiv.org/abs/2204.03762}{{\tt
  arXiv:2204.03762 [astro-ph.GA]}}.

\bibitem{Gomes:2023xzk}
D.~A. Gomes, R.~Briffa, A.~Kozak, J.~Levi~Said, M.~Saal, and A.~Wojnar,
  ``{Cosmological constraints of Palatini f(\ensuremath{\mathscr{R}})
  gravity},'' \href{http://dx.doi.org/10.1088/1475-7516/2024/01/011}{{\em JCAP}
  {\bf 01} (2024)  011}, \href{http://arxiv.org/abs/2310.17339}{{\tt
  arXiv:2310.17339 [gr-qc]}}.

\bibitem{jimenez2015tensor}
J.~Beltran~Jimenez, L.~Heisenberg, and G.~J. Olmo, ``{Tensor perturbations in a
  general class of Palatini theories},''
  \href{http://dx.doi.org/10.1088/1475-7516/2015/06/026}{{\em JCAP} {\bf 06}
  (2015)  026}, \href{http://arxiv.org/abs/1504.00295}{{\tt arXiv:1504.00295
  [gr-qc]}}.

\bibitem{jimenez2017gravitational}
J.~Beltran~Jimenez, L.~Heisenberg, G.~J. Olmo, and D.~Rubiera-Garcia, ``{On
  gravitational waves in Born-Infeld inspired non-singular cosmologies},''
  \href{http://dx.doi.org/10.1088/1475-7516/2017/10/029}{{\em JCAP} {\bf 10}
  (2017)  029}, \href{http://arxiv.org/abs/1707.08953}{{\tt arXiv:1707.08953
  [hep-th]}}. [Erratum: JCAP 08, E01 (2018)].

\bibitem{jimenez2018born}
J.~Beltran~Jimenez, L.~Heisenberg, G.~J. Olmo, and D.~Rubiera-Garcia,
  ``{Born\textendash{}Infeld inspired modifications of gravity},''
  \href{http://dx.doi.org/10.1016/j.physrep.2017.11.001}{{\em Phys. Rept.} {\bf
  727} (2018)  1--129}, \href{http://arxiv.org/abs/1704.03351}{{\tt
  arXiv:1704.03351 [gr-qc]}}.

\bibitem{Maggiore:2007ulw}
M.~Maggiore,
  \href{http://dx.doi.org/10.1093/acprof:oso/9780198570745.001.0001}{{\em
  {Gravitational Waves. Vol. 1: Theory and Experiments}}}.
\newblock Oxford University Press, 2007.

\bibitem{Garoffolo:2019mna}
A.~Garoffolo, G.~Tasinato, C.~Carbone, D.~Bertacca, and S.~Matarrese,
  ``{Gravitational waves and geometrical optics in scalar-tensor theories},''
  \href{http://dx.doi.org/10.1088/1475-7516/2020/11/040}{{\em JCAP} {\bf 11}
  (2020)  040}, \href{http://arxiv.org/abs/1912.08093}{{\tt arXiv:1912.08093
  [gr-qc]}}.

\bibitem{fleurry_thesis}
P.~Fleury, ``{Light propagation in inhomogeneous and anisotropic
  cosmologies},'' \href{http://arxiv.org/abs/1511.03702}{{\tt arXiv:1511.03702
  [gr-qc]}}.

\bibitem{Deffayet:2007kf}
C.~Deffayet and K.~Menou, ``{Probing Gravity with Spacetime Sirens},''
  \href{http://dx.doi.org/10.1086/522931}{{\em Astrophys. J. Lett.} {\bf 668}
  (2007)  L143--L146}, \href{http://arxiv.org/abs/0709.0003}{{\tt
  arXiv:0709.0003 [astro-ph]}}.

\bibitem{Saltas:2014dha}
I.~D. Saltas, I.~Sawicki, L.~Amendola, and M.~Kunz, ``{Anisotropic Stress as a
  Signature of Nonstandard Propagation of Gravitational Waves},''
  \href{http://dx.doi.org/10.1103/PhysRevLett.113.191101}{{\em Phys. Rev.
  Lett.} {\bf 113} (2014) no.~19, 191101},
  \href{http://arxiv.org/abs/1406.7139}{{\tt arXiv:1406.7139 [astro-ph.CO]}}.

\bibitem{Nishizawa:2017nef}
A.~Nishizawa, ``{Generalized framework for testing gravity with
  gravitational-wave propagation. I. Formulation},''
  \href{http://dx.doi.org/10.1103/PhysRevD.97.104037}{{\em Phys. Rev. D} {\bf
  97} (2018) no.~10, 104037}, \href{http://arxiv.org/abs/1710.04825}{{\tt
  arXiv:1710.04825 [gr-qc]}}.

\bibitem{Belgacem:2017ihm}
E.~Belgacem, Y.~Dirian, S.~Foffa, and M.~Maggiore, ``{Gravitational-wave
  luminosity distance in modified gravity theories},''
  \href{http://dx.doi.org/10.1103/PhysRevD.97.104066}{{\em Phys. Rev. D} {\bf
  97} (2018) no.~10, 104066}, \href{http://arxiv.org/abs/1712.08108}{{\tt
  arXiv:1712.08108 [astro-ph.CO]}}.

\bibitem{Newman:1961qr}
E.~Newman and R.~Penrose, ``{An Approach to gravitational radiation by a method
  of spin coefficients},'' \href{http://dx.doi.org/10.1063/1.1724257}{{\em J.
  Math. Phys.} {\bf 3} (1962)  566--578}.

\bibitem{Chandrasekhar:1985kt}
S.~Chandrasekhar, {\em {The mathematical theory of black holes}}.
\newblock Clarendon Press, 1985.

\bibitem{Fluerry_2015}
P.~Fleury, C.~Pitrou, and J.-P. Uzan, ``{Light propagation in a homogeneous and
  anisotropic universe},''
  \href{http://dx.doi.org/10.1103/PhysRevD.91.043511}{{\em Phys. Rev. D} {\bf
  91} (2015) no.~4, 043511}, \href{http://arxiv.org/abs/1410.8473}{{\tt
  arXiv:1410.8473 [gr-qc]}}.

\bibitem{Tasinato:2021wol}
G.~Tasinato, A.~Garoffolo, D.~Bertacca, and S.~Matarrese, ``{Gravitational-wave
  cosmological distances in scalar-tensor theories of gravity},''
  \href{http://dx.doi.org/10.1088/1475-7516/2021/06/050}{{\em JCAP} {\bf 06}
  (2021)  050}, \href{http://arxiv.org/abs/2103.00155}{{\tt arXiv:2103.00155
  [gr-qc]}}.

\bibitem{Sasaki:1987ad}
M.~Sasaki, ``{The Magnitude - Redshift relation in a perturbed Friedmann
  universe},'' {\em Mon. Not. Roy. Astron. Soc.} {\bf 228} (1987)  653--669.

\bibitem{Navarro:1996gj}
J.~F. Navarro, C.~S. Frenk, and S.~D.~M. White, ``{A Universal density profile
  from hierarchical clustering},'' \href{http://dx.doi.org/10.1086/304888}{{\em
  Astrophys. J.} {\bf 490} (1997)  493--508},
  \href{http://arxiv.org/abs/astro-ph/9611107}{{\tt arXiv:astro-ph/9611107}}.

\bibitem{Ettori:2018tus}
S.~Ettori, V.~Ghirardini, D.~Eckert, E.~Pointecouteau, F.~Gastaldello,
  M.~Sereno, M.~Gaspari, S.~Ghizzardi, M.~Roncarelli, and M.~Rossetti,
  ``{Hydrostatic mass profiles in X-COP galaxy clusters},''
  \href{http://dx.doi.org/10.1051/0004-6361/201833323}{{\em Astron. Astrophys.}
  {\bf 621} (2019)  A39}, \href{http://arxiv.org/abs/1805.00035}{{\tt
  arXiv:1805.00035 [astro-ph.CO]}}.

\bibitem{Meneghetti}
M.~{Meneghetti}, \href{http://dx.doi.org/10.1007/978-3-030-73582-1}{{\em
  {Introduction to Gravitational Lensing; With Python Examples}}}, vol.~956.
\newblock 2021.

\bibitem{Petters}
A.~O. {Petters}, H.~{Levine}, and J.~{Wambsganss}, {\em {Singularity theory and
  gravitational lensing}}.
\newblock 2001.

\bibitem{Treu}
T.~Treu, ``Strong lensing by galaxies,''
  \href{http://dx.doi.org/10.1146/annurev-astro-081309-130924}{{\em Annual
  Review of Astronomy and Astrophysics} {\bf 48} (2010) no.~1, 87--125}.
  \url{https://doi.org/10.1146/annurev-astro-081309-130924}.

\bibitem{Kormann1994}
R.~Kormann, P.~Schneider, and M.~Bartelmann, ``{Isothermal elliptical
  gravitational lens models},'' {\em Astronomy and Astrophysics} {\bf 284}
  (1994)  285--299.

\end{thebibliography}\endgroup

\appendix

\section{Sachs formalism}\label{appendix_B}

With the notations of Sec.(\ref{In GR}), 
let us introduce an orthonormal basis (the Sachs basis) $(s^{\mu}_{A})$, $A = \{1,2\}$, satisfying the conditions
\begin{equation}
    s^{\mu}_{A}u_{\mu} =    s^{\mu}_{A}d_{\mu} = 0,  \qquad  s^{\mu}_{A}s_{B\mu} = \delta_{AB}.
\end{equation}
In the further part, we follow %the notation from 
\cite{fleurry_thesis}. 

One of the ways to study the beam pattern is provided by the solutions of the deviation equation \eqref{deviation} provided in the Sachs basis
\begin{equation}\label{sachs}
    \frac{d^{2}\chi^{A}}{d\lambda^{2}} = \mathcal{R}^{A}_{B}\chi^{B}
\end{equation}
where $\mathcal{R}^{A}_{B}$ is the optical tidal matrix which is symmetric under the exchange of their indices. Moreover, it can be decomposed into pure trace and trace-free parts as 
\begin{equation}    
\mathcal{R} = 
\begin{pmatrix}
    \mathscr{R}& 0\\ 0& \mathscr{R}
\end{pmatrix} + \begin{pmatrix} -\text{Re}\mathscr{W}& \text{Im}\mathscr{W}\\ \text{Im}\mathscr{W}&\text{Re}\mathscr{W}
\end{pmatrix}
\end{equation}
where $\mathscr{R}=-\frac{1}{2}R_{\mu\nu}k^{\mu}k^{\nu}$  is the Ricci lensing term, responsible for a  homothetic transformation of the beam pattern. On the other hand, 
\begin{equation}
    \mathscr{W} = -\frac{1}{2}C_{\mu\rho\nu\sigma}(s^{\mu}_{1} -i s^{\mu}_{2})k^{\rho}k^{\nu}(s^{\sigma}_{1} -i s^{\sigma}_{2}),
\end{equation}
 is the Weyl lensing matrix responsible for the elongation and contraction of the beam pattern in different directions. The Weyl tensor $C_{\mu\rho\nu\sigma}$ has the same symmetries as the Riemann tensor and it is trace-free . As we will see, these quantities are useful in lensing studies to express the wave's polarization amplitude. % as they influence the geometry of the beam in different ways

Let us now denote the optical tidal matrix as  $\mathcal{R}$. %$ = \mathcal{R}_{AB}$.
Then, the Sachs equation \eqref{sachs} can be written in terms of the Jacobi matrix $\mathcal{D}$ which relates the position of images on the observers' celestial sphere to the physical separation of the sources:
\begin{equation}\label{jacobi_mat_eqn}
    \ddot{\mathcal{D}} = \mathcal{R}\mathcal{D},
\end{equation}
where dot denotes the derivative with respect to the affine parameter $\lambda$. Let us now define the deformation rate of the beam as
\begin{equation}\label{deformation_matrix}
    \mathcal{S} = \frac{\dot{\mathcal{D}}}{\mathcal{D}}.
\end{equation}
Note that the deformation matrix \eqref{deformation_matrix} can be expressed in terms of Sach's basis as
\begin{equation}\label{Sbasis}
\mathcal{S}_{AB} =  s^{\mu}_{A}s^{\nu}_{B}\nabla_{\mu}k_{\nu}.
\end{equation}
Now, we can decompose the matrix $\mathcal{S}$ into anti-symmetric, trace, and traceless symmetric parts
\begin{equation}\label{matrixs}
    \mathcal{S} =\begin{pmatrix} 0 & \omega\\-\omega & 0\end{pmatrix} +\begin{pmatrix} \theta & 0 \\ 0& \theta \end{pmatrix} + \begin{pmatrix} -\sigma_{1} & \sigma_{2} \\ \sigma_{2} & \sigma_{1} \end{pmatrix},
\end{equation}
where $\omega$, $\theta$ and $\sigma = \sigma_{1} + i\sigma_{2}$ are optical scalars defined as
\begin{equation}
    \omega^{2}=\frac{1}{2}\omega^{\alpha\beta}\omega_{\alpha\beta}\,, \;\;
    \theta = \frac{1}{2}k^\alpha_{\,;\alpha} \,, \;\;
    \sigma^{2} = \sigma^{\alpha\beta}\sigma_{\alpha\beta}
\end{equation}
with 
\begin{align}
    \omega_{\alpha\beta} &= k_{[\alpha;\beta]} \\
    \sigma_{\alpha\beta} &= k_{(\alpha;\beta)}-\frac{1}{2}Q_{\alpha\beta}\theta \,.
\end{align}
The semicolon $_{;\beta}$ means the application of $\nabla_\beta$ while the round and rectangular brackets denote symmetrization and antisymmetrization in the indices, respectively.
Since we are dealing with a null congruence, the vorticity tensor $ \omega_{\alpha\beta}$ is zero since the null vector was defined as $k_{\mu} = \Phi,_{\mu}$ after the expression \eqref{Eansatz}. Therefore, the matrix's components \eqref{matrixs} consist of the expansion $\theta$ and shear scalars $\sigma_1$ and $\sigma_2$ only.

\section{Weak-field limit equations for Palatini gravity}\label{appA}

In the weak-field limit, the metric in modified gravity can be written as 
\begin{equation}\label{Pmetric}
    ds^{2} =  -(1+2\Phi )dt^{2} + (1-2\Psi)(dx^{2} + dy^{2}+ dz^{2}),
\end{equation}
where $\Phi$ and $\Psi$ are Bardeen potentials. We can directly obtain the non-vanishing components of the affine connection corresponding to the background geometry associated with the lensing mass:
\begin{align}
    \delta \hat{\Gamma}^{0}_{i0} &= -\partial_{i}\Phi  + \Phi \partial_{i} f', \\
\delta \hat{\Gamma}^{i}_{00} &=  \partial_{i}\Psi, \\
\delta  \hat{\Gamma}^{0}_{ij} &= 0 ,\\
\delta \hat{\Gamma}^{i}_{j0} &= 0 ,\\    
 \delta \Gamma ^{i}_{jk} &= - \partial_{k}\Psi \delta^{i}_{j} - \delta^{i}_{k}
 \partial_{j}\Psi + \delta_{jk}\partial^{i}\Psi \nonumber\\  &-\partial_{j} \ln f' \delta^{i}_{k}\Psi -\partial_{k} \ln f' \delta^{i}_{j}\Psi +\partial^{i} \ln f' \delta_{jk}\Psi.
\end{align}
\section{Luminosity distance}\label{luminosity_distance}

A key observable in gravitational wave  domain is the luminosity distance. It is defined in terms of the ratio of GW power emitted by the source and flux registered at the detector location. Following the treatment of \cite{Sasaki:1987ad,Tasinato:2021wol}, the gravitational energy flux measured by an observer can be calculated according to the following formula:
\begin{equation}
    \hat{\mathcal{F}}^{\alpha} = - \hat{T}^{\mu}_{\nu}h^{\alpha}_{\mu}u^{\nu} 
\end{equation}
where $\hat{T}_{\mu\nu}$ is the gravitational wave energy momentum tensor in Palatini $f(\hat{R})$ gravity, $h^{\alpha}_{\mu}$  is the projection tensor and $u^{\alpha}$ the four-velocity of the observer. The projection tensor can be written as 
$h^{\alpha}_{\mu} = \delta^{\alpha}_{\mu} + u^{\alpha}u_{\mu}$ and energy-momentum tensor can be obtained from  the amplitude evolution equation following \cite{Tasinato:2021wol}, in which we assume that the structure of the $\hat{T}_{\mu\nu}$ is the same as in GR with modified amplitude derived in \eqref{P_amp_evolution}
\begin{equation}
    \hat{T}_{\mu\nu} = \frac{1}{32\pi}C^{P}\hat{k}_{\mu}\hat{k}_{\nu},
\end{equation}
where $C^{P} = C^{GR}e^{-\int(2\frac{d \ln f'}{d\lambda} + \frac{1}{2}f'\frac{d\ln f'}{d\lambda} )d\lambda }$ is the evolution of amplitude given in eqn.\eqref{P_amp_evolution}, and $\hat{k}^{\mu}$ is the wave vector in the conformal frame.  Using this one obtains the measured energy flux as 

\begin{equation}
   \hat{  \mathcal{F}} = \mathcal{F}\hat{d} = \frac{1}{32\pi}(C^{P})^2 \omega^2 \hat{d},
\end{equation}
where $\mathcal{F}$ is the flux amplitude, and $\hat{d}^{\alpha}$ is the unit space-like vector given by
\begin{equation}
    \hat{d}^{\alpha} = \frac{1}{\omega}(\hat{k}^{\alpha} - \omega \hat{u}^{\alpha}),
\end{equation}
where $\omega$ is the gravitational wave frequency measured by the observer, defined as $\omega = - k^{\mu}u_{\mu}$. This notion underlies the definition of the GW redshift for a given value of the affine parameter $\lambda$, as: 
\begin{equation}
    1 + z(\lambda) = \frac{\omega(\lambda)}{\omega(0)}.
\end{equation}
If we assume that GWs are emitted by a spherically symmetric source with radius $R_{s}$, then the intrinsic luminosity of the source is given by
$L_{s} = 4\pi R_{s}^2 \mathcal{F}(\lambda_{s})$, where as usual $\lambda_s$ is the value of the affine parameter at the source. The luminosity distance to the source as measured by observed is given by 
\begin{equation}
    d_{L} = \Big[ \frac{L_s}{4\pi \mathcal{F}(\lambda_{O})}\Big]^{1/2} = \Big[ \frac{\mathcal{F}(\lambda_s)}{\mathcal{F}(\lambda_{O})}\Big]^{1/2} R_{s}.
\end{equation}
This gives us the luminosity distance relation 

\begin{equation}\label{luminosity_distance_corr1}
    d^{\text{P}}_{L} = d^{\text{GR}}_{L}\exp \Big[ \int_{\lambda_{s}}^{\lambda_{o}}\Big(2 \frac{d \ln f'}{d \lambda}+\frac{1}{2} f' \frac{d \ln f'}{d\lambda} \Big)d\lambda \Big].
\end{equation}
It is to be noted that in lensing angular diameter distances are used and we assume Etherington's
reciprocity theorem. However its validity in the chosen theory of gravity need to be proved.

\section{lensing potential and convergence} \label{appLensingJacobian}
First-order strong lensing effects, i.e. image distortion and magnification can be derived from the Jacobi matrix of lens equation:
\begin{equation}
    [A]_{ij} := \left(\frac{\partial \vec{y}}{\partial \vec{x}} \right)_{ij} = \left(\delta_{ij} - \frac{\partial^2 \Psi}{\partial x_i \partial x_j} \right).
\end{equation}
With a shorthand notation $\Psi_{ij}$ for partial derivatives in the above expression, one can express the shear tensor 
$\Gamma = \begin{pmatrix}
    \gamma_1 & \gamma_2 \\
    \gamma_2 & - \gamma_1
\end{pmatrix}
$ 
where: $\gamma_1 = \frac{1}{2}\left( \Psi_{11} - \Psi_{22}\right) =: \gamma \cos{2 \varphi} $, $\gamma_2 = \Psi_{12} = \Psi_{21}=: \gamma \sin{2 \varphi} $ For the SIS lens in Palatini theory, one has 
\begin{eqnarray}
    \Psi_{11} &=& \frac{1}{x^5} \left[ x^2 \left( x^2 - \frac{2 \alpha}{\xi_0^2} \right) + \left( \frac{6 \alpha}{\xi_0^2} - x^2 \right) x_1^2 \right], \nonumber \\
    \Psi_{12} &=& \Psi_{21} =  \frac{x_1 x_2}{x^5} \left( \frac{6 \alpha}{\xi_0^2} - x^2 \right)  \nonumber, \\
    \Psi_{22} &=& \frac{1}{x^5} \left[ x^2 \left( x^2 - \frac{2 \alpha}{\xi_0^2} \right) + \left( \frac{6 \alpha}{\xi_0^2} - x^2 \right) x_2^2 \right] \nonumber \,,
\end{eqnarray}
hence the shear components are
\begin{eqnarray}
    \gamma_1 &=& \frac{1}{2 x^5} \left( \frac{6 \alpha}{\xi_0^2} - x^2 \right) \left(x_1^2 - x_2^2 \right) ,\nonumber \\
    \gamma_2 &=& \left( \frac{6 \alpha}{\xi_0^2} - x^2 \right) \frac{x_1 x_2}{x^5} \,,
\end{eqnarray}
where the notation $\vec{x} = (x_1,x_2)$ and $x = \sqrt{x_1^2 + x_2^2}$ was used.

%\begin{equation}
%    F = - \nabla \Phi(r) = 4\pi G( \ln r - \frac{2\alpha}{r^{2}})
%\end{equation}

%\begin{eqnarray}
%     &&\Psi(\Vec{x}) = \frac{1}{\pi}\int_{\textbf{R}^{2}}\kappa(\vec{x'})(\ln|\vec{x} - \vec{x'}|  + (\vec{x} - \vec{x'}) %\nonumber\\
%     &&(\ln|\vec{x} - \vec{x'}|) -(\vec{x} - \vec{x'})) + \frac{2\alpha}{\vec{x} - \vec{x'}} )d^{2}x'.
%\end{eqnarray}

\end{document}